\documentclass[journal]{IEEEtran}
\usepackage{epsfig,graphics,subfigure,epsf,color,amsmath,balance,cite,amsfonts,multirow,stfloats,algorithm,algorithmic,mathrsfs,amssymb}
\usepackage{mathtools}

\newtheorem{theorem}{Theorem}

\newtheorem{remark}{Remark}
\newtheorem{proposition}{Proposition}

\title{Beamforming Optimization for IRS-Aided Communications with Transceiver Hardware Impairments}

\author{Hong Shen,~\IEEEmembership{Member,~IEEE,}~Wei Xu,~\IEEEmembership{Senior Member,~IEEE,}~Shulei Gong,\\ Chunming Zhao,~\IEEEmembership{Member,~IEEE,}~and~Derrick Wing Kwan Ng,~\IEEEmembership{Senior Member,~IEEE}\thanks{The work of H. Shen and W. Xu was supported by the National Natural Science Foundation of China under Grants 61871108, 62022026, and 61871109, and the Natural Science Foundation of Jiangsu Province under Grant BK20190012. The work of D. W. K. Ng was supported by funding from the UNSW Digital Grid Futures Institute, UNSW, Sydney, under a cross-disciplinary fund scheme and by the Australian Research Council's Discovery Project (DP190101363). \emph{(Corresponding authors: Wei Xu.)}}\thanks{H. Shen and C. Zhao are with the National Mobile Communications Research Laboratory, Southeast University, Nanjing 210096, China (e-mail: shhseu@seu.edu.cn; cmzhao@seu.edu.cn).} \thanks{W. Xu is with the National Mobile Communications Research Laboratory, Southeast University, Nanjing 210096, China, and is also with the Purple Mountain Laboratories, Nanjing 211111, China (email: wxu@seu.edu.cn).} \thanks{S. Gong is with the China Mobile Group Jiangsu Co., Ltd, Nanjing, 210029, China (email: gongshulei@smail.nju.edu.cn).} \thanks{D. W. K. Ng is with the School of Electrical Engineering and Telecommunications, University of New South Wales, NSW 2052, Australia (email: w.k.ng@unsw.edu.au).}}

\begin{document}

\maketitle

\begin{abstract}
In this paper, we focus on intelligent reflecting surface (IRS) assisted multi-antenna communications with transceiver hardware impairments encountered in practice. In particular, we aim to maximize the received signal-to-noise ratio (SNR) taking into account the impact of hardware impairments, where the source transmit beamforming and the IRS reflect beamforming are jointly designed under the proposed optimization framework. To circumvent the non-convexity of the formulated design problem, we first derive a closed-form optimal solution to the source transmit beamforming. Then, for the optimization of IRS reflect beamforming, we obtain an upper bound to the optimal objective value via solving a single convex problem. A low-complexity minorization-maximization (MM) algorithm was developed to approach the upper bound.  Simulation results demonstrate that the proposed beamforming design is more robust to the hardware impairments than that of the conventional SNR maximized scheme. Moreover, compared to the scenario without deploying an IRS, the performance gain brought by incorporating the hardware impairments is more evident for the IRS-aided communications.
\end{abstract}


\section{Introduction}
With the rapid development of electromagnetic materials, intelligent reflecting surface (IRS) has become a cost-effective promising solution for future wireless communications {\cite{Liaskos2018CM,Wu2020CM,Basar2019ArxivCM,Renzo2019Eura,Zhang2019Arxiv,Chen2020Arxiv,Renzo2020JSAC}}.  By deploying a number of passive reflecting elements with controllable phase shifts, IRS has the capability of improving communication quality via enhancing received signal power and harnessing interference power at desired receivers, {which paves the way to realizing smart radio environments \cite{Renzo2019Eura,Renzo2020JSAC}}. Compared to the well-known massive multiple-input multiple-output (MIMO) technique {\cite{Marzetta2010TWC,Rusek2013SPM}}, {IRS can provide excellent system performance} without requiring power-hungry components such as power amplifiers\footnote{IRS is also different from the large intelligent surface (LIS) recently proposed in \cite{Hu2018TSP} due to the fact that IRS is a passive device while LIS is an active one.}.
As a result, resource allocation design for IRS-assisted systems has drawn significant research interest recently. For example, the transmission optimization for various IRS-assisted communication systems has been studied in a number of recent works \cite{Wu2018GLOBECCOM,Wu2019ICASSP,Wu2019TWC,Wu2020TCOM,Huang2018ICASSP,Guo2019Arxiv,Huang2019TWC,Huang2018GLOBECCOM,Hong2019CL,Yu2019GLOBECOM,Cui2019WCL,Guan2020WCL,Yu2020JSAC,Wu2019ArxivWCL}. Specifically, the received signal-to-noise ratio (SNR) maximization \cite{Wu2018GLOBECCOM,Wu2019ICASSP} and  quality-of-service (QoS) constrained transmit power minimization \cite{Wu2019TWC,Wu2020TCOM} problems were investigated for an IRS-aided single-user multiple-input
single-output (MISO) system. Regarding a multi-user MISO system assisted by an IRS, different design objectives, e.g., transmit power minimization  \cite{Wu2019TWC,Wu2020TCOM}, achievable rate maximization \cite{Huang2018ICASSP,Guo2019Arxiv}, and energy efficiency maximization \cite{Huang2019TWC,Huang2018GLOBECCOM}, have been adopted for optimizing the beamforming scheme.  In addition to  single-user and multi-user systems, beamforming optimization for  IRS-assisted multi-antenna wiretap channels was investigated in \cite{Hong2019CL,Yu2019GLOBECOM,Cui2019WCL,Guan2020WCL,Yu2020JSAC}, where the design goal is to maximize the system achievable secrecy rate.  {Alternatively, focusing on an IRS-aided multi-antenna  simultaneous wireless information and power transfer (SWIPT) system, the authors of \cite{Wu2019ArxivWCL} performed beamforming optimization to maximize the weighted sum power received by energy harvesting receivers while fulfilling the QoS constraints for information decoding receivers.} As shown in all the above works, the presence of an IRS complicates the optimal beamforming design  due to the fact that the phase shifts of an IRS are not only coupled with the transmit beamforming, but they should also satisfy unit-modulus constraints or belong to a discrete set \cite{Wu2019ICASSP,Wu2020TCOM,Guo2019Arxiv,Huang2018GLOBECCOM}.

For practical communication systems, there exist inevitable impairments in transceiver hardware  such as amplifiers, oscillators, digital-to-analog converters (DACs), and analog-to-digital converters (ADCs) {\cite{Suzuki2008JSAC,Namgoong2005TWC}}. Note that different from the traditional noise at the receiver side, hardware impairments lead to distortions on both transmitted and received signals. More importantly, the power of distortions scales with the power of the useful signal at the wireless transceivers. Hence, several efforts have been made to analyze the impact of hardware impairments on various MIMO systems \cite{Bjornson2013CL,Bjornson2014TIT,Zhu2017TWC}. As a remedy, the authors of \cite{Bjornson2012GLOBECOM,You2017CL,Shen2017Access,Taghizadeh2018TWC} conducted transceiver optimization for MIMO systems such as multicell MIMO and MIMO relaying, in order to relieve the performance deterioration caused by the hardware impairments. {Concerning the IRS-aided multi-antenna communications, the existence of transceiver hardware impairments, if not carefully handled, will affect both the source transmit beamforming and the IRS reflect beamforming designs, thus leading to severer performance degradation than that in conventional systems without exploiting an IRS.} Hence, it is necessary to take into account the hardware impairments when  performing  beamforming optimization for IRS-aided multi-antenna systems, which,  to the best of our knowledge, has not been studied in literature yet.

In this paper, motivated by the aforementioned observations, we study the beamforming optimization for an IRS-aided multi-antenna communication system by incorporating signal distortions caused by hardware impairments. Both source transmit beamforming and IRS reflect beamforming are optimized to maximize the received SNR at the destination. Compared with a relevant work \cite{Wu2018GLOBECCOM} under the assumption of perfect hardware, the considered problem in this paper is more sophisticated due to the presence of hardware impairments and cannot be directly solved with the methods developed in \cite{Wu2018GLOBECCOM}. {On the other hand, existing robust designs, e.g., \cite{Bjornson2012GLOBECOM,You2017CL,Shen2017Access,Taghizadeh2018TWC}, for conventional multi-antenna systems are not applicable to IRS-assisted systems since the IRS phase shifts also need to be optimized, which is non-trivial even for the perfect hardware case.} Our main contributions are summarized as follows:
\begin{itemize}
\item
We determine the optimal source transmit beamforming, which has a closed form and is in fact a weighted version of matched filter (MF) beamforming.
\item We acquire an upper bound to the optimal objective value of the IRS reflect beamforming optimization, which can be efficiently calculated by solving one convex problem.
\item We further employ a low-complexity minorization-maximization (MM) technique to handle the IRS reflect beamforming optimization. The method only requires calculating a closed-form expression iteratively while almost achieving the upper bound.
\end{itemize}

\emph{Notations:} Vectors and matrices are represented by boldface lower-case and boldface upper-case letters, respectively. $|\cdot|$ and $\|\cdot\|$ stand for the absolute value of the input scalar and the Euclidean norm of the input vector, respectively. $(\cdot)^*$, $(\cdot)^{T}$, and $(\cdot)^{H}$ denote the conjugate, the transpose, and the Hermitian transpose operations, respectively. $\text{arg}(\cdot)$  and $\Re(\cdot)$ return the phase and the real part of the input complex number, respectively. $\text{diag}\{\cdot\}$ represents a diagonal matrix with diagonal elements being the entries of the input vector  and $ {\widetilde {\text{diag}}}\{\cdot\}$ denotes a diagonal matrix which has the same diagonal elements as the input matrix. $(\cdot)^{-1}$, $\text{tr}(\cdot)$, and $\lambda_\text{max}(\cdot)$ denote the  inverse, the trace, and the maximum eigenvalue of the input matrix, respectively.  $(\cdot)_{i,j}$ denotes the ($i$, $j$)-th entry of the input matrix. $\mathbf a(1:N)$ denotes the vector whose entries are the first $N$ entries of vector $\mathbf a$. $a \sim \mathcal{CN}(\mu,\sigma^2)$ means that scalar $a$ is a circularly symmetric complex Gaussian random variable with mean $\mu$ and variance $\sigma^2$. $\mathbf a \sim \mathcal{CN}(\boldsymbol \mu,\mathbf \Sigma)$ means that vector $\mathbf a$ follows circularly symmetric complex Gaussian distribution with mean $\boldsymbol \mu$ and covariance matrix $\mathbf \Sigma$.  $\mathbf A \succeq \mathbf 0$ denotes that matrix $\mathbf A$ is positive semidefinite.  $\mathbb{E}\{\cdot\}$ is the expectation operation. $\mathbb{C}^{m \times n}$ stands for the set of $m \times n$ complex matrices. $\mathbf e_m$ is the unit vector with the $m$-th element being 1. $\mathbf I_n$ denotes the $n \times n$ identity matrix.  $O(\cdot)$ represents the big-O notation.

\begin{figure}[t]
    \begin{center}
      \epsfxsize=7.0in\includegraphics[scale=0.6]{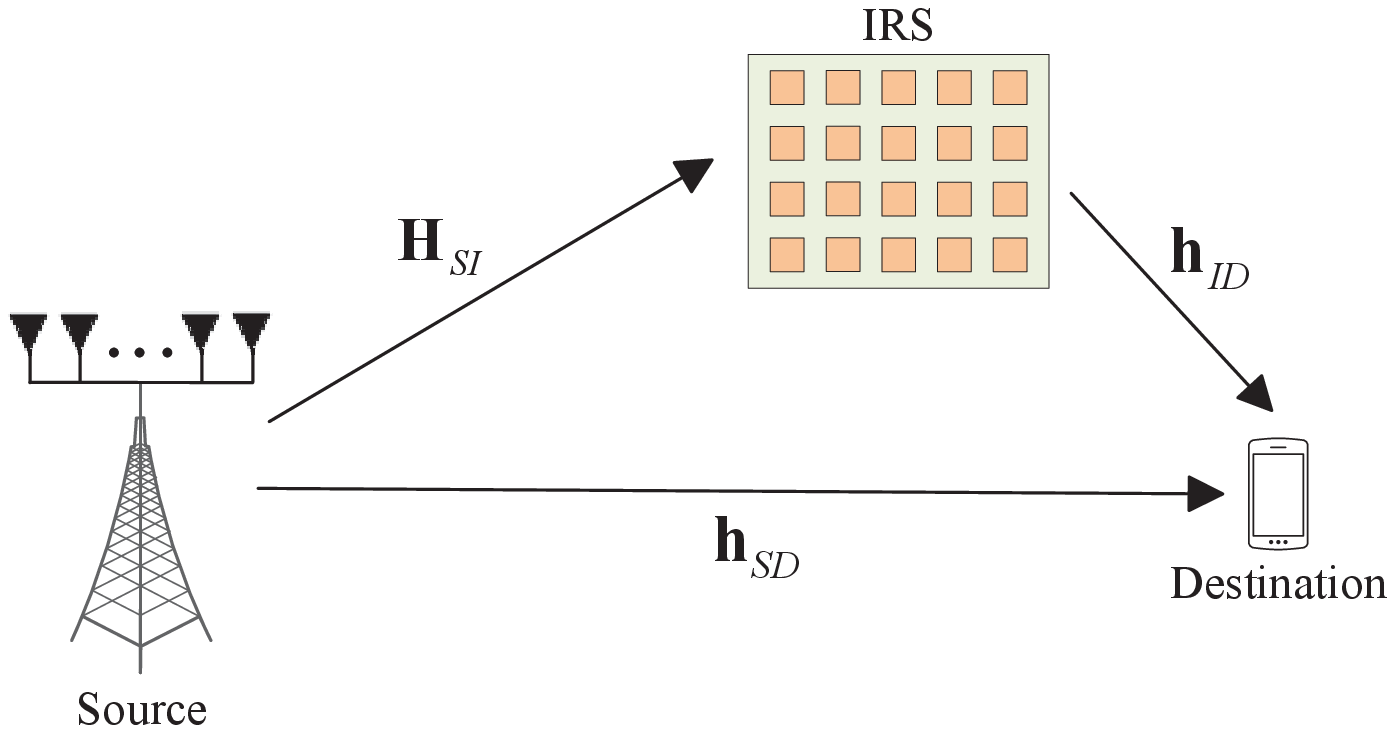}
      \caption{{A wireless communication system assisted by an IRS.}}\label{fig:model}
    \end{center}
\end{figure}

\begin{figure*}[ht]
\normalsize
\newcounter{mytempeqncnt0}
\setcounter{mytempeqncnt0}{\value{equation}}
\setcounter{equation}{2}
\begin{align}\label{eq:SNR}
\text{SNR}(\mathbf{w}, \mathbf{\Theta})= \frac{|(\mathbf{h}_{ID}^H\mathbf{\Theta}\mathbf{H}_{SI}+
\mathbf{h}_{SD}^H)\mathbf w|^2}{(\mathbf{h}_{ID}^H\mathbf{\Theta}\mathbf{H}_{SI}+
\mathbf{h}_{SD}^H)(\kappa_D\mathbf w\mathbf w^H+(1+\kappa_D)\kappa_S{\widetilde {\text{diag}}}\{\mathbf w\mathbf w^H\})(\mathbf{H}_{SI}^H\mathbf{\Theta}^H\mathbf{h}_{ID}+
\mathbf{h}_{SD})+(1+\kappa_D)\sigma_{n}^2}
\end{align}
\setcounter{equation}{\value{mytempeqncnt0}}
\vspace*{4pt}
\hrulefill
\end{figure*}
\begin{figure*}[ht]
\normalsize
\newcounter{mytempeqncnt1}
\setcounter{mytempeqncnt1}{\value{equation}}
\setcounter{equation}{4}
\begin{align}\label{eq:SNRMaxProbw}
\mathop{\text{maximize}}\limits_{\mathbf{w}} \quad & \frac{|(\mathbf{h}_{ID}^H\mathbf{\Theta}\mathbf{H}_{SI}+
\mathbf{h}_{SD}^H)\mathbf w|^2}{(\mathbf{h}_{ID}^H\mathbf{\Theta}\mathbf{H}_{SI}+
\mathbf{h}_{SD}^H)(\kappa_D\mathbf w\mathbf w^H+(1+\kappa_D)\kappa_S{\widetilde {\text{diag}}}\{\mathbf w\mathbf w^H\})(\mathbf{H}_{SI}^H\mathbf{\Theta}^H\mathbf{h}_{ID}+
\mathbf{h}_{SD})+(1+\kappa_D)\sigma_{n}^2} \nonumber\\
\text{subject to}\quad  & \|\mathbf{w}\|^2 \leq \tilde P
\end{align}
\setcounter{equation}{\value{mytempeqncnt1}}
\vspace*{4pt}
\hrulefill
\end{figure*}
\begin{figure*}[ht]
\normalsize
\newcounter{mytempeqncnt2}
\setcounter{mytempeqncnt2}{\value{equation}}
\setcounter{equation}{5}
\begin{align}\label{eq:SNRMaxProbw1}
\mathop{\text{maximize}}\limits_{\mathbf{w}} \quad & \frac{|(\mathbf{h}_{ID}^H\mathbf{\Theta}\mathbf{H}_{SI}+
\mathbf{h}_{SD}^H)\mathbf w|^2}{\mathbf{w}^H(\kappa_D\mathbf \Upsilon(\mathbf{\Theta})+(1+\kappa_D)\kappa_S{\widetilde {\text{diag}}}\{\mathbf \Upsilon(\mathbf{\Theta})\})\mathbf{w}+(1+\kappa_D)\sigma_{n}^2} \nonumber\\
\text{subject to}\quad\  & \|\mathbf{w}\|^2 \leq \tilde P
\end{align}
\setcounter{equation}{\value{mytempeqncnt2}}
\vspace*{4pt}
\hrulefill
\end{figure*}
\section{System Model and Problem Formulation}\label{sec:model}
\subsection{System Model}\label{sec:modelA}
We consider an IRS-aided communication system consisting of one source, one IRS, and one destination, as depicted in Fig.~\ref{fig:model}. The source is equipped with $N_S$ antennas, the IRS has $N_I$  passive reflecting elements, and the destination is a single-antenna device.

Different from prior works on IRS-aided communications, e.g., \cite{Wu2018GLOBECCOM,Wu2019ICASSP,Wu2019TWC,Wu2020TCOM,Huang2018ICASSP,Guo2019Arxiv,Huang2019TWC,Huang2018GLOBECCOM,Hong2019CL,Yu2019GLOBECOM,Cui2019WCL,Guan2020WCL,Yu2020JSAC,Wu2019ArxivWCL}, we incorporate practical transceiver hardware impairments in the considered system. Specifically, the source transmit signal is expressed by
\begin{align}\label{eq:txsignal}
\mathbf t=\mathbf w x+\mathbf z_S,
\end{align}
where $\mathbf w \in \mathbb{C}^{N_S \times 1}$ is the source transmit beamforming, $x$ is the transmit symbol of the source with unit power, and
{$\mathbf z_S$ represents the transmit distortion noise independent of $x$, which models the joint effects of the power amplifier noise, the non-linearities
in DAC and power amplifier, and the oscillator phase noise \cite{Suzuki2008JSAC}. In particular, the entries of $\mathbf z_S$ are independent zero-mean Gaussian random variables and the variance of the $i$-th entry is proportional to the power of the intended transmit signal at the $i$-th antenna, i.e., $\mathbf z_S\sim \mathcal{CN}(\mathbf 0,\kappa_S {\widetilde {\text{diag}}}\{\mathbf w\mathbf w^H\})$, where $\kappa_S \in (0,1)$ denotes the normalized variance of the transmit distortion noise.}

The received signal at the destination is expressed by
\begin{align}\label{eq:recsignal}
{y}=&\ \underbrace{\mathbf{h}_{ID}^H\mathbf{\Theta}\mathbf{H}_{SI}\mathbf t}_{\text{reflected link}}+
\underbrace{\mathbf{h}_{SD}^H\mathbf t}_{\text{direct link}}+{n+z_D}\nonumber\\
\triangleq &\ \tilde y+z_D,
\end{align}
where $\mathbf{h}_{ID} \in \mathbb{C}^{N_I \times 1}$ is the IRS-to-destination channel, $\mathbf{\Theta}\triangleq\text{diag}\{[e^{j\phi_1},\cdots,e^{j\phi_i},\cdots,e^{j\phi_{N_I}}]\}$ is the reflect beamforming matrix of the IRS with $\phi_i$ denoting the phase shift incurred by the $i$-th reflecting element, $\mathbf{H}_{SI}  \in \mathbb{C}^{N_I \times N_S}$ is the source-to-IRS channel,  $\mathbf{h}_{SD}  \in \mathbb{C}^{N_S \times 1}$ is the source-to-destination channel, and ${n}\sim \mathcal{CN}(0,\sigma_{n}^2)$ is the additive white Gaussian noise (AWGN) at the destination. {Variable $z_D$ stands for the receive distortion noise independent of $\tilde y$, which models the joint effects of the automatic gain control (AGC) noise, the non-linearities in ADC and AGC, and the oscillator phase noise \cite{Namgoong2005TWC}. Note that $z_D$ follows zero-mean Gaussian distribution and its variance is proportional to the power of the undistorted received signal, i.e.,  $z_D \sim \mathcal{CN}(0,\kappa_D\mathbb{E}\{|\tilde y|^2\})$, where $\kappa_D \in (0,1) $ denotes the normalized variance of the receive distortion noise.}


Based on \eqref{eq:txsignal} and \eqref{eq:recsignal}, we obtain the received SNR at the destination which is given by \eqref{eq:SNR}.
Since the source transmit beamformer $\mathbf w$ appears in both the numerator and the denominator of the above SNR expression, it is expected that the received SNR of the considered system in the presence of transceiver hardware impairments will become saturated when the transmit power is exceedingly large.

{We note that the reflecting element of the IRS is subject to the phase errors \cite{Godara1985JOE,Peng2014Conf} caused by quantized phase shifts in practice. This is because the IRS usually has a large number of reflecting elements and the implementation cost per reflecting element should be as low as possible to reduce the overall cost.
In fact, the reflecting element of the IRS typically has a small number of discrete phase shifts, e.g., 2 or 4 \cite{Tan2018Conf,Dai2020Access,Xu2020AWPL}, which results in non-negligible phase errors. To capture the impacts of IRS phase errors in the current framework, we can simply replace the ideal continuous phase shift constraints by discrete constraints as in \cite{Wu2019ICASSP,Guo2019Arxiv,Huang2018GLOBECCOM}, where the discrete phase shift set is given by  $\mathcal{F}_d \triangleq \left\{\phi_i \mid \phi_i \in \left\{0,\frac{2\pi}{L},\cdots,\frac{2\pi(L-1)}{L}\right\}\right\}$ with $L=2^B$ being the number of phase shift levels and $B$ denoting the number of bits used to represent the $L$ phase shift levels.}


\subsection{Problem Formulation}
In this paper, we aim to maximize the received SNR in \eqref{eq:SNR} by jointly optimizing the source transmit beamformer $\mathbf w$ and the IRS reflect beamforming matrix  $\mathbf{\Theta}$, where we also impose a power constraint on $\mathbf w$ and unit-modulus constraints on the diagonal elements of $\mathbf{\Theta}$. More specifically, according to \eqref{eq:txsignal} and the distribution of $\mathbf z_S$, we can readily derive the total transmit power of the source node as $(1+\kappa_S)\|\mathbf w\|^2$. {Moreover, both continuous and discrete phase shift constraints will be considered in this work, where the result of the continuous phase shift case serves as a performance benchmark.} Therefore, we formulate the design problem as
\setcounter{equation}{3}
\begin{align}\label{eq:SNRMaxProb}
\mathop{\text{maximize}}\limits_{\mathbf{w},\mathbf{\Theta}} \quad &
\text{SNR}(\mathbf{w}, \mathbf{\Theta})\nonumber\\
\text{subject to}\quad &  (1+\kappa_S)\|\mathbf{w}\|^2 \leq P,\nonumber\\ &{\theta_i \in \mathcal{F}_1 \text{ or }\mathcal{F}_2},\ i=1,\cdots,N_I,
\end{align}
where $P$ denotes the maximal source transmit power and $\theta_i$ is the $i$-th diagonal element of $\mathbf{\Theta}$. {$\mathcal{F}_1$ and $\mathcal{F}_2$ are defined by $\mathcal{F}_1 \triangleq \left\{\theta_i\mid |\theta_i|=1\right\}$ and $\mathcal{F}_2 \triangleq \left\{\theta_i\mid\theta_i=e^{j\phi_i},\phi_i \in \left\{0,\frac{2\pi}{L},\cdots,\frac{2\pi(L-1)}{L}\right\}\right\}$, respectively.}

The above problem has a complicated objective function and nonconvex unit-modulus constraints, whose optimal solution cannot be readily determined. In particular, compared to the SNR maximization problem in \cite{Wu2018GLOBECCOM}, the presence of hardware impairments makes the beamforming optimization more challenging and thus existing methods designed for the case of perfect hardware are not applicable to the problem at hand.


\begin{figure*}[ht]
\normalsize
\newcounter{mytempeqncnt3}
\setcounter{mytempeqncnt3}{\value{equation}}
\setcounter{equation}{6}
\begin{align}\label{eq:wopt}
\mathbf{w}^\star=\sqrt{\tilde P}\frac{\left(\kappa_D\mathbf \Upsilon(\mathbf{\Theta})+(1+\kappa_D)\kappa_S{\widetilde {\text{diag}}}\{\mathbf \Upsilon(\mathbf{\Theta})\}+\frac{(1+\kappa_D)\sigma_{n}^2}{\tilde P}\mathbf I_{N_S}\right)^{-1}(\mathbf{H}_{SI}^H\mathbf{\Theta}^H\mathbf{h}_{ID}+
\mathbf{h}_{SD})}{\left\|\left(\kappa_D\mathbf \Upsilon(\mathbf{\Theta})+(1+\kappa_D)\kappa_S{\widetilde {\text{diag}}}\{\mathbf \Upsilon(\mathbf{\Theta})\}+\frac{(1+\kappa_D)\sigma_{n}^2}{\tilde P}\mathbf I_{N_S}\right)^{-1}(\mathbf{H}_{SI}^H\mathbf{\Theta}^H\mathbf{h}_{ID}+
\mathbf{h}_{SD})\right\|}
\end{align}
\setcounter{equation}{\value{mytempeqncnt3}}
\vspace*{4pt}
\hrulefill
\end{figure*}
\begin{figure*}[ht]
\normalsize
\newcounter{mytempeqncnt4}
\setcounter{mytempeqncnt4}{\value{equation}}
\setcounter{equation}{8}
\begin{align}\label{eq:SNRMaxProbwtheta}
\mathop{\text{maximize}}\limits_{\mathbf{\Theta}}  \quad &  \tilde P(\mathbf{h}_{ID}^H\mathbf{\Theta}\mathbf{H}_{SI}+
\mathbf{h}_{SD}^H)\Bigg(\kappa_D\mathbf \Upsilon(\mathbf{\Theta})+(1+\kappa_D)\kappa_S   {\widetilde {\text{diag}}}\{\mathbf \Upsilon(\mathbf{\Theta})\}+\frac{(1+\kappa_D)\sigma_{n}^2}{\tilde P}\mathbf I_{N_S}\Bigg)^{-1} (\mathbf{H}_{SI}^H\mathbf{\Theta}^H\mathbf{h}_{ID}\!+\!
\mathbf{h}_{SD})\nonumber\\
\text{subject to}\quad & |\theta_i|=1,\ i=1,\cdots,N_I
\end{align}
\setcounter{equation}{\value{mytempeqncnt4}}
\vspace*{4pt}
\hrulefill
\end{figure*}
\section{Closed-Form Optimal Solution to Source Transmit Beamforming}
In this section, we determine a closed-form optimal solution to $\mathbf w$  for an arbitrary $\mathbf{\Theta}$. Based on problem \eqref{eq:SNRMaxProb}, the optimization of $\mathbf w$ with a fixed $\mathbf{\Theta}$ is expressed by \eqref{eq:SNRMaxProbw},
where $\tilde P\triangleq\frac{P}{1+\kappa_S}$. To find the optimal solution to this problem in a closed form, we need to first transform the denominator of the objective function into a tractable form. In fact, by invoking the equality $\mathbf a^H {\widetilde {\text{diag}}}\{\mathbf b\mathbf b^H\}\mathbf a=\mathbf b^H {\widetilde {\text{diag}}}\{\mathbf a\mathbf a^H\}\mathbf b$, we transform problem \eqref{eq:SNRMaxProbw} into an equivalent form in \eqref{eq:SNRMaxProbw1},
where $\mathbf \Upsilon(\mathbf{\Theta})\triangleq(\mathbf{H}_{SI}^H\mathbf{\Theta}^H\mathbf{h}_{ID}+
\mathbf{h}_{SD})(\mathbf{h}_{ID}^H\mathbf{\Theta}\mathbf{H}_{SI}+
\mathbf{h}_{SD}^H)$.
{By dividing the numerator and the denominator of the above objective function by $\|\mathbf{w}\|^2$, we can readily confirm that it is monotonically increasing with $\|\mathbf{w}\|$.} Hence, the power consumption constraint must be active at optimality. Moreover, by utilizing the generalized Rayleigh quotient \cite{Golub1996}, we obtain the optimal solution to $\mathbf{w}$ in \eqref{eq:wopt}.

Note that a matrix inversion is required by \eqref{eq:wopt}, which can be computationally inefficient particularly when $N_S$ is large. To deal with this issue, we show that $\mathbf{w}^\star$ has an equivalent form that only involves the inversion of a diagonal matrix in the following proposition.
\begin{proposition}\label{prop:wopt}
The optimal solution to problem \eqref{eq:SNRMaxProbw1} can be expressed by
\setcounter{equation}{7}
\begin{align}\label{eq:wopt1}
\mathbf{w}^\star=\sqrt{\tilde P}\frac{\mathbf {\tilde \Upsilon}^{-1}(\mathbf{\Theta})(\mathbf{H}_{SI}^H\mathbf{\Theta}^H\mathbf{h}_{ID}+
\mathbf{h}_{SD})}{\left\|\mathbf {\tilde \Upsilon}^{-1}(\mathbf{\Theta})(\mathbf{H}_{SI}^H\mathbf{\Theta}^H\mathbf{h}_{ID}+
\mathbf{h}_{SD})\right\|},
\end{align}
where $\mathbf {\tilde \Upsilon}(\mathbf{\Theta})\triangleq (1+\kappa_D)\kappa_S{\widetilde {\text{diag}}}\{\mathbf \Upsilon(\mathbf{\Theta})\}+\frac{(1+\kappa_D)\sigma_{n}^2}{\tilde P}\mathbf I_{N_S}$.
\end{proposition}
\begin{IEEEproof}
See Appendix~\ref{app:wopt}.
\end{IEEEproof}

Compared to \eqref{eq:wopt}, we only need to calculate the inversion of a diagonal matrix, thus requiring much lower computational complexity. Moreover, it can be observed from \eqref{eq:wopt1} that $\mathbf{w}^\star$ is actually a product of a diagonal weighted matrix and the MF beamforming, where the factor of hardware impairments is included in the weighted matrix. {When there exist severe hardware impairments, i.e., $\kappa_S$ or $\kappa_D$ is large, the weights become smaller such that the performance degradation caused by hardware impairments is alleviated. On the other hand, for the ideal scenario without transceiver hardware impairments, i.e., $\kappa_S=\kappa_D=0$, it is clear that \eqref{eq:wopt1} reduces to the conventional MF beamforming.}

\section{Optimization of IRS Reflect Beamforming}\label{sec:thetaopt}
{Now, let us first consider the IRS reflect beamforming optimization under the continuous phase shift constraints, i.e., $|\theta_i|=1,\ i=1,\cdots,N_I$.} Recall that the optimal transmit beamforming was obtained in a closed form given any reflect beamforming in the previous section.  By plugging the optimal $\mathbf{w}^\star$ in \eqref{eq:wopt} into the original problem in \eqref{eq:SNRMaxProb}, we transform problem \eqref{eq:SNRMaxProb} into \eqref{eq:SNRMaxProbwtheta} without loss of optimality.
Denote the above objective function by $\psi(\mathbf{\Theta})$. Then, by left multiplying $\mathbf{h}_{ID}^H\mathbf{\Theta}\mathbf{H}_{SI}+
\mathbf{h}_{SD}^H$ on both sides of  \eqref{eq:woptinv1} in Appendix~\ref{app:wopt}, we have
\setcounter{equation}{9}
\begin{align}\label{eq:psitheta}
\psi(\mathbf{\Theta})=\frac{\tilde P\tilde \psi(\mathbf{\Theta})}{\kappa_D \tilde \psi(\mathbf{\Theta})+1},
\end{align}
where $\tilde \psi(\mathbf{\Theta})\triangleq(\mathbf{h}_{ID}^H\mathbf{\Theta}\mathbf{H}_{SI}+
\mathbf{h}_{SD}^H)\mathbf {\tilde \Upsilon}^{-1}(\mathbf{\Theta}) (\mathbf{H}_{SI}^H\mathbf{\Theta}^H\mathbf{h}_{ID}+
\mathbf{h}_{SD})$.
Hence, we are able to convert problem \eqref{eq:SNRMaxProbwtheta} to the simplified form as follows:
\begin{align}\label{eq:SNRMaxProbwtheta0}
\mathop{\text{maximize}}\limits_{\mathbf{\Theta}}  \ &  (\mathbf{h}_{ID}^H\mathbf{\Theta}\mathbf{H}_{SI}\!+\!
\mathbf{h}_{SD}^H)\mathbf {\tilde \Upsilon}^{-1}(\mathbf{\Theta}) (\mathbf{H}_{SI}^H\mathbf{\Theta}^H\mathbf{h}_{ID}\!+\!
\mathbf{h}_{SD})\nonumber\\
\text{subject to}\ & |\theta_i|=1,\ i=1,\cdots,N_I.
\end{align}
Different from problem \eqref{eq:SNRMaxProbw} with respect to $\mathbf{w}$, it is quite difficult to optimally solve this problem  using the generalized Rayleigh quotient due to the more complicated objective function and nonconvex constraints.  In the following, we first obtain an upper bound to the optimal objective value of problem \eqref{eq:SNRMaxProbwtheta0}. Then, we propose an MM-based algorithm to approach this upper bound with low computational complexity. {Finally, based on the obtained continuous solution, we provide a method to achieve an efficient solution for the discrete phase shift case.}

\subsection{Upper Bound to Optimal Objective Value}\label{sec:UpperBnd}
Although it is challenging to achieve the globally optimal solution to problem \eqref{eq:SNRMaxProbwtheta0} in polynomial time due to its nonconvexity, we are able to acquire an upper bound to the optimal objective value via solving a convex problem as given in the subsequent theorem.
\begin{theorem}\label{theo:upper}
The optimal objective value of problem \eqref{eq:SNRMaxProbwtheta0} is upper bounded by that of the following convex problem:
\begin{align}\label{eq:SNRMaxProbwtheta5}
\mathop{\text{maximize}}\limits_{\boldsymbol {\tilde \Theta} \succeq \mathbf 0}  \quad &  \sum_{m=1}^{N_S} \frac{(\mathbf \Psi \boldsymbol {\tilde \Theta}\mathbf \Psi^H)_{m,m}}{(1+\kappa_D)\kappa_S(\mathbf \Psi \boldsymbol {\tilde \Theta}\mathbf \Psi^H)_{m,m}+\frac{(1+\kappa_D)\sigma_{n}^2}{\tilde P}}\nonumber\\
\text{subject to}\quad & (\boldsymbol {\tilde \Theta})_{i,i}=1,\ i=1,\cdots,N_I+1,
\end{align}
where $\mathbf \Psi\triangleq[\mathbf{H}_{SI}^H\text{diag}\{\mathbf{h}_{ID}\}\ \mathbf{h}_{SD}]$ {and $\boldsymbol {\tilde \Theta}$ is a positive semidefinite matrix of size $N_I+1$}.
\end{theorem}
\begin{IEEEproof}
{See Appendix~\ref{app:theoupper}.}
\end{IEEEproof}
According to the above conclusion, we can determine an upper bound to the optimal objective value of problem \eqref{eq:SNRMaxProbwtheta0} by solving problem \eqref{eq:SNRMaxProbwtheta5} with the off-the-shelf convex optimization tools such as CVX \cite{Grant2018CVX}. This upper bound serves as a benchmark to evaluate the performance of our proposed algorithm.

\begin{figure*}[ht]
\normalsize
\newcounter{mytempeqncnt5}
\setcounter{mytempeqncnt5}{\value{equation}}
\setcounter{equation}{12}
\begin{align}\label{eq:fbndfinal}
& \boldsymbol {\tilde \theta}^H\mathbf \Psi^H\left((1+\kappa_D)\kappa_S{\widetilde {\text{diag}}}\{\mathbf \Psi \boldsymbol {\tilde \theta}\boldsymbol {\tilde \theta}^H\mathbf \Psi^H\}+\frac{(1+\kappa_D)\sigma_{n}^2}{\tilde P}\mathbf I_{N_S}\right)^{-1} \mathbf \Psi \boldsymbol {\tilde \theta} \nonumber \\
\geq & 2\Re\{\boldsymbol {\tilde \theta}_0^H(\mathbf \Psi^H \mathbf \Xi_0^{-1}\mathbf \Psi -(1+\kappa_D)\kappa_S(\mathbf \Omega-\lambda_\text{max}(\mathbf \Omega)\mathbf I_{N_I+1})) \boldsymbol {\tilde \theta}\}-2(1+\kappa_D)\kappa_S(N_I+1)\lambda_\text{max}(\mathbf \Omega)\nonumber \\
& +\boldsymbol {\tilde \theta}_0^H(2(1+\kappa_D)\kappa_S\mathbf \Omega-\mathbf \Psi^H \mathbf \Xi_0^{-1}\mathbf \Psi)\boldsymbol {\tilde \theta}_0
\end{align}
\setcounter{equation}{\value{mytempeqncnt5}}
\vspace*{4pt}
\hrulefill
\end{figure*}
\begin{figure*}[ht]
\normalsize
\newcounter{mytempeqncnt6}
\setcounter{mytempeqncnt6}{\value{equation}}
\setcounter{equation}{13}
\begin{align}\label{eq:SNRMaxProbwthetaMM}
\mathop{\text{maximize}}\limits_{\boldsymbol {\tilde \theta}}  \quad &  \Re\left\{\left(\boldsymbol {\tilde \theta}^{(k-1)}\right)^H(\mathbf \Psi^H \mathbf \Xi_0^{-1}\mathbf \Psi -(1+\kappa_D)\kappa_S(\mathbf \Omega-\lambda_\text{max}(\mathbf \Omega)\mathbf I_{N_I+1})) \boldsymbol {\tilde \theta}\right\}\nonumber\\
\text{subject to}\quad & |\tilde \theta_i|=1,\ i=1,\cdots,N_I+1
\end{align}
\setcounter{equation}{\value{mytempeqncnt6}}
\vspace*{4pt}
\hrulefill
\end{figure*}
\subsection{MM Algorithm for Solving Problem \eqref{eq:SNRMaxProbwtheta0}}\label{sec:MMalg}
We consider solving problem \eqref{eq:SNRMaxProbwtheta0}, or, equivalently, problem \eqref{eq:SNRMaxProbwtheta3} in Appendix~\ref{app:theoupper}, via the popular MM algorithm \cite{Sun2017TSP}, which achieves almost the same objective value as the upper bound obtained in Section~\ref{sec:UpperBnd} with low computational cost.

According to the philosophy of the MM technique, we first construct a surrogate function that serves as a lower bound of the objective function of problem \eqref{eq:SNRMaxProbwtheta3}, which is shown in the following proposition.
\begin{proposition}\label{prop:MMbnd}
For problem \eqref{eq:SNRMaxProbwtheta3} in Appendix~\ref{app:theoupper}, a lower bound of the objective function used for the MM algorithm is given by \eqref{eq:fbndfinal},
where $\mathbf \Xi_0\triangleq(1+\kappa_D)\kappa_S{\widetilde {\text{diag}}}\{\mathbf \Psi \boldsymbol {\tilde \theta}_0\boldsymbol {\tilde \theta}_0^H\mathbf \Psi^H\}+\frac{(1+\kappa_D)\sigma_{n}^2}{\tilde P}\mathbf I_{N_S}$ with $\boldsymbol {\tilde \theta}_0$ fulfilling the unit-modulus constraints and $\mathbf \Omega\triangleq\mathbf \Psi^H {\widetilde {\text{diag}}}\{\mathbf \Xi_0^{-1}\mathbf \Psi  \boldsymbol {\tilde \theta}_0  \boldsymbol {\tilde \theta}_0^H\mathbf \Psi^H\mathbf \Xi_0^{-1}\} \mathbf \Psi$. 
\end{proposition}
\begin{IEEEproof}
See Appendix~\ref{app:MMbnd}.
\end{IEEEproof}

\begin{algorithm}[t]
\caption{\textbf{MM Algorithm for Problem \eqref{eq:SNRMaxProbwtheta0}}}\label{alg2}

\begin{algorithmic}[1]

\STATE \textit{Initialization:}  Initialize $\boldsymbol {\tilde \theta}^{(0)}$, iteration index $k=0$, and convergence accuracy $\epsilon \rightarrow 0$.
\STATE \textbf{repeat}

\STATE \text{  } $k=k+1$.

\STATE \text{  }  {Calculate $\boldsymbol {\tilde \theta}^{\star}$ using \eqref{eq:thetaiopt}.}

\STATE \text{  }  {$\boldsymbol {\tilde \theta}^{(k)}=\boldsymbol {\tilde \theta}^{\star}$.}

\STATE \textbf{until} convergence.

\STATE Calculate $\boldsymbol {\Theta} ^ \star =\text{diag}\{(\boldsymbol {\tilde \theta}^{(k)}(1:N_I)/{\tilde \theta}_{N_I+1}^{(k)})^*\}$.
\end{algorithmic}
\end{algorithm}
Then, the MM algorithm is adopted to iteratively tighten the lower bound of the objective function. Specifically, in the $k$-th iteration of the MM algorithm, we set $\boldsymbol {\tilde \theta}_0=\boldsymbol {\tilde \theta}^{(k-1)}$ with $\boldsymbol {\tilde \theta}^{(k-1)}$ denoting the optimal solution to $\boldsymbol {\tilde \theta}$ obtained in the $(k-1)$-th iteration, and consider maximizing the lower bound in \eqref{eq:fbndfinal}. Accordingly, the corresponding problem to be solved in the $k$-th iteration is given by \eqref{eq:SNRMaxProbwthetaMM},
where we remove the constant terms in \eqref{eq:fbndfinal}. Define $\boldsymbol \alpha\triangleq (\mathbf \Psi^H \mathbf \Xi_0^{-1}\mathbf \Psi -(1+\kappa_D)\kappa_S(\mathbf \Omega-\lambda_\text{max}(\mathbf \Omega) \mathbf I_{N_I+1}))\boldsymbol {\tilde \theta}^{(k-1)}$. Then, we can readily obtain the optimal solution to problem \eqref{eq:SNRMaxProbwthetaMM} by
\setcounter{equation}{14}
\begin{align}\label{eq:thetaiopt}
\tilde \theta_i^{\star}=e^{j\text{arg}(\alpha_i)},\ i=1,\cdots,N_I+1,
\end{align}
where $\alpha_i$ is the $i$-th entry of $\boldsymbol \alpha$.

The proposed MM algorithm for solving problem \eqref{eq:SNRMaxProbwtheta0} is summarized in Algorithm~\ref{alg2}, where the last step follows from the definition of $\boldsymbol {\tilde \theta}$ in Appendix~\ref{app:theoupper}. Regarding the convergence property of Algorithm~\ref{alg2}, we can prove the following conclusion.
\begin{theorem}\label{theo:converge}
Algorithm~\ref{alg2} converges to a stationary point of problem \eqref{eq:SNRMaxProbwtheta0} when convergence is reached.
\end{theorem}
\begin{IEEEproof}
See Appendix~\ref{app:theo}.
\end{IEEEproof}

After obtaining $\boldsymbol {\Theta} ^ \star$, i.e., the solution to IRS reflect beamforming, via Algorithm~\ref{alg2}, we substitute it into \eqref{eq:wopt1} to calculate $\mathbf w ^ \star$, i.e., the solution to source transmit beamforming. {We would like to emphasize that the proposed solution framework does not optimize the source transmit beamforming and the IRS reflect beamforming in an alternating manner as the one commonly applied in the literature, e.g., \cite{Hong2019CL,Yu2019GLOBECOM,Cui2019WCL,Guan2020WCL,Yu2020JSAC}. Moreover, based on Theorem~\ref{theo:converge} and the equivalence between problem \eqref{eq:SNRMaxProbwtheta0} and the original beamforming optimization problem in \eqref{eq:SNRMaxProb}, we arrive at the conclusion that $(\mathbf{w}^\star,\mathbf{\Theta}^\star)$ is a stationary point of problem \eqref{eq:SNRMaxProb}.}

{Based on the continuous solution to the IRS phase shifts obtained by Algorithm~\ref{alg2}, we provide a simple but effective method to acquire discrete phase shifts. Specifically, we project the IRS phase shifts $\phi_i^\star,\ i=1,\cdots,N_I,$ calculated by Algorithm~\ref{alg2} onto the discrete set $\mathcal{F}_d$ defined in Section~\ref{sec:modelA}, i.e., $\hat \phi_i=\text{arg}\mathop{\text{min}}\limits_{\phi \in \mathcal{F}_d}|\phi_i^\star-\phi|,\ i=1,\cdots,N_I$, where $\hat \phi_i$ is the quantized phase shift of the $i$-th reflecting element. As will be shown in Section~\ref{sec:simu}, we can achieve a performance approaching that of the continuous phase shift case by setting only $B=2$.}

\begin{figure*}[ht]
\normalsize
\newcounter{mytempeqncnt7}
\setcounter{mytempeqncnt7}{\value{equation}}
\setcounter{equation}{15}
\begin{align}\label{eq:SNRMaxProbwtheta0DL}
{\mathop{\text{maximize}}\limits_{\mathbf{\Theta}}}  \quad  &  {\mathbf{h}_{ID}^H\mathbf{\Theta}\mathbf{H}_{SI}\left(\!(1\!+\!\kappa_D)\kappa_S{\widetilde {\text{diag}}}\{\mathbf{H}_{SI}^H\mathbf{\Theta}^H\mathbf{h}_{ID}\mathbf{h}_{ID}^H\mathbf{\Theta}\mathbf{H}_{SI}\}\!+\!\frac{(1\!+\!\kappa_D)\sigma_{n}^2}{\tilde P}\mathbf I_{N_S}\!\right)^{-1} \mathbf{H}_{SI}^H\mathbf{\Theta}^H\mathbf{h}_{ID}}\nonumber\\
{\text{subject to}} \quad & {|\theta_i|=1,\ i=1,\cdots,N_I}
\end{align}
\setcounter{equation}{\value{mytempeqncnt7}}
\vspace*{4pt}
\hrulefill
\end{figure*}
\begin{figure*}[ht]
\normalsize
\newcounter{mytempeqncnt8}
\setcounter{mytempeqncnt8}{\value{equation}}
\setcounter{equation}{16}
\begin{align}\label{eq:obj}
& {|\eta|^2|\mathbf{h}_{ID}^H\mathbf{\Theta}\mathbf a_I|^2\mathbf a_S^H\left((1+\kappa_D)\kappa_S{\widetilde {\text{diag}}}\{|\eta|^2|\mathbf{h}_{ID}^H\mathbf{\Theta}\mathbf a_I|^2\mathbf a_S\mathbf a_S^H\}+\frac{(1+\kappa_D)\sigma_{n}^2}{\tilde P}\mathbf I_{N_S}\right)^{-1}\mathbf a_S}\nonumber \\
 {=}& {\frac{N_S|\eta|^2|\mathbf{h}_{ID}^H\mathbf{\Theta}\mathbf a_I|^2}{(1+\kappa_D)\kappa_S|\eta|^2|\mathbf{h}_{ID}^H\mathbf{\Theta}\mathbf a_I|^2+\frac{(1+\kappa_D)\sigma_{n}^2}{\tilde P}}}
\end{align}
\setcounter{equation}{\value{mytempeqncnt8}}
\vspace*{4pt}
\hrulefill
\end{figure*}
\begin{remark}
We find empirically that the convergence of Algorithm~\ref{alg2}  can be further accelerated  by applying the acceleration scheme based on SQUAREM \cite[Section V-B]{Song2016TSP}. In fact, from the comparison results in Table~\ref{table:iteration} of Section~\ref{sec:simu}, using the acceleration method significantly reduces the required average number of iterations.
\end{remark}

\begin{remark}
{For the proposed design, the optimized solutions to the source transmit beamforming and the IRS reflect beamforming  can be calculated at the source with acceptable computational complexity (see Section~\ref{sec:comp} for details). Also, the calculated IRS discrete phase shifts (represented by bits) are conveyed to the IRS via a dedicated wireless control link as in prior works \cite{Wu2019TWC,Wu2020TCOM}. On the other hand, the beamforming calculation at the source requires the normalized distortion noise variances $\kappa_S$ and $\kappa_D$. The two constants can be measured at the source and the destination in an offline manner, respectively, and $\kappa_D$ can be fed back to the source for beamforming calculation. In addition, channel state information (CSI) is also needed for beamforming calculation. Fortunately, we find that it suffices to acquire $\mathbf {\Psi} \triangleq [ \mathbf H_{SI}^H \text{diag}\{\mathbf h_{ID} \} \  \mathbf h_{SD}]$ for the calculation of the proposed beamforming solution. In fact, by properly setting the IRS phase shifts during the channel training, the source can directly estimate $\mathbf {\Psi}$ with some conventional channel estimation methods such as least-square (LS) channel estimation \cite{Jensen2020Conf,You2020Conf}.}
\end{remark}

\begin{remark}
{The proposed beamforming design is also applicable to a time division multiple access (TDMA)-based multi-user system. For this scenario, we only need to apply the proposed method to each user in each independent time slot. Nevertheless, we recognize that the extension to other multi-user systems, e.g., spatial division multiple access (SDMA)-based multi-user system, is non-trivial. In fact, the beamforming optimization method developed in our work may no longer be optimal due to the more complicated objective function. The study of the beamforming optimization for the SDMA-based multi-user system with hardware impairments is left for future works.}
\end{remark}

{\subsection{Special Case Without Direct Link}}
{In practice, the direct link between the source and the destination is negligible if there exists an exceedingly large path loss between them. By neglecting the direct link, i.e., $\mathbf h_{SD}^H=\mathbf 0$, the IRS reflect beamforming optimization problem in \eqref{eq:SNRMaxProbwtheta0} becomes \eqref{eq:SNRMaxProbwtheta0DL}.}
{In order to obtain a simple and insightful solution to the above problem, we further assume that the source-to-IRS channel $\mathbf{H}_{SI}$ only has the line-of-sight (LOS) component, i.e., $\mathbf{H}_{SI}=\eta\mathbf a_I\mathbf a_S^H$, where $\eta$ is the complex channel gain, and $\mathbf a_I$ and $\mathbf a_S$, with normalized entries, are the array response vectors of the IRS and the source, respectively. Then, the objective function of problem \eqref{eq:SNRMaxProbwtheta0DL} is rewritten by \eqref{eq:obj}}
{since $\|\mathbf a_S\|^2=N_S$ and $\widetilde {\text{diag}}\{\mathbf a_S\mathbf a_S^H\}=\mathbf I_{N_S}$. Therefore, it follows that problem \eqref{eq:SNRMaxProbwtheta0DL} can be simplified to}
\setcounter{equation}{17}
\begin{align}\label{eq:SNRMaxProbwtheta1DL}
{\mathop{\text{maximize}}\limits_{\mathbf{\Theta}} } \quad & {|\mathbf{h}_{ID}^H\mathbf{\Theta}\mathbf a_I|} \nonumber\\
{\text{subject to}}\quad & {|\theta_i|=1,\ i=1,\cdots,N_I.}
\end{align}
{According to \cite[Section IV]{Wu2018GLOBECCOM}, the optimal solution to the above problem is given by}
\begin{align}\label{eq:thetaiDL}
{\theta_i^\star=e^{-j(\text{arg}(h_{ID,i}^*)+\text{arg}(a_{I,i}))},\ i=1,\cdots,N_I,}
\end{align}
{where $h_{ID,i}$ is the $i$-th entry of $\mathbf{h}_{ID}$ and $a_{I,i}$ is the $i$-th entry of $\mathbf a_I$. Moreover, by substituting \eqref{eq:thetaiDL} into the optimal source transmit beamforming in \eqref{eq:wopt1}, we can readily obtain}
\begin{align}\label{eq:woptDL}
{\mathbf w^\star=\sqrt{\frac{\tilde P}{N_S}}\mathbf a_S.}
\end{align}
{Therefore, when the direct link is negligible and the source-to-IRS channel only contains the LOS component, the optimal source transmit beamforming should be aligned with the source array response vector while the optimal IRS phase shifts should match the sum of the phase shifts of the IRS-to-destination channel and the IRS array response vector.}

{By substituting \eqref{eq:thetaiDL} and \eqref{eq:woptDL} into the expression of the received SNR in \eqref{eq:SNR}, we have}
\begin{align}
{\text{SNR}=  \frac{ \tilde PN_S|\eta|^2\|\mathbf{h}_{ID}\|_1^2}{\tilde P(\kappa_DN_S+(1+\kappa_D)\kappa_S)|\eta|^2\|\mathbf{h}_{ID}\|_1^2+{(1+\kappa_D)\sigma_{n}^2}}.}
\end{align}
{To gain further insights, we assume that $\mathbf{h}_{ID} \sim \mathcal{CN}(\mathbf 0,\sigma_{ID}^2\mathbf I_{N_I})$, where $\sigma_{ID}^2$ denotes the variance of each entry of $\mathbf{h}_{ID}$. Then, $|{h}_{ID,i}|$ follows Rayleigh distribution with mean $\sqrt{\pi}\sigma_{ID}/2$. Accordingly, when $N_I \rightarrow \infty$, $\frac{\|\mathbf{h}_{ID}\|_1}{N_I} \rightarrow \sqrt{\pi}\sigma_{ID}/2$ and thus we obtain the asymptotic SNR as}
\begin{align}\label{eq:SNRinf}
{\text{SNR} \rightarrow  \frac{ \tilde PN_S|\eta|^2\pi N_I^2\sigma_{ID}^2}{\tilde P(\kappa_DN_S\!+\!(1\!+\!\kappa_D)\kappa_S)|\eta|^2\pi N_I^2\sigma_{ID}^2\!+\!{4(1\!+\!\kappa_D)\sigma_{n}^2}}.}
\end{align}
{It can be observed from \eqref{eq:SNRinf} that when the destination noise variance $\sigma_n^2$ is non-negligible, the received SNR is monotonically increasing with respect to $N_I$ and $\sigma_{ID}^2$. Hence, we can compensate the performance degradation caused by hardware impairments by increasing the size of the IRS and/or reducing the distance between the IRS and the destination. On the other hand, when the destination noise is negligible, the received SNR is less sensitive to the size of the IRS or the distance between the IRS and the destination in the large limit of $N_I$.}

\section{Simulation Results and Complexity Analysis}\label{sec:simu}
\subsection{Simulation Setup}\label{subsec:simuA}
In this section, we perform simulations to compare the performance of the following beamforming designs:
\begin{itemize}
\item Proposed robust design incorporating hardware impairments for IRS-aided systems (``Robust w/ IRS'');
\item Nonrobust design neglecting hardware impairments for IRS-aided systems (``Nonrobust w/ IRS'');
\item Robust design incorporating hardware impairments for conventional systems without IRS (``Robust w/o IRS'');
\item Nonrobust design neglecting hardware impairments for conventional systems without IRS (``Nonrobust w/o IRS'').
\end{itemize}
For the reflect beamforming optimization of ``Robust w/ IRS'', we generate an initial point $\boldsymbol {\tilde \theta}^{(0)}$ by normalizing each entry of a random $(N_I+1) \times 1$ vector. ``Nonrobust w/ IRS'' is obtained by setting $\kappa_S=\kappa_D=0$ for ``Robust w/ IRS''. ``Robust w/o IRS'' is obtained by setting $\mathbf \Theta=\mathbf 0$ in \eqref{eq:wopt1}  and ``Nonrobust w/o IRS'' is achieved by applying MF beamforming at the source pointing towards the desired receiver.

\begin{table*}[t]
\renewcommand{\arraystretch}{1.3}
\caption{Simulation Parameters}
\label{table:simupara}
\centering
\begin{tabular}{|c|c|}
\hline
Parameter Name & Parameter Value\\
\hline
Number of Source Transmit Antennas $N_S$ & $4$ \\
\hline
Number of Reflecting Elements $N_I$ & $50$ \\
\hline
Maximum Source Transmit Power $P$ & $12$ dBW\\
\hline
Normalized Distortion Power $\kappa=\kappa_S=\kappa_D$ & $0.07$\\
\hline
Destination Noise Power $\sigma_n^2$ & $-85$ dBW\\
\hline
Reference Distance $d_0$ &
$1$ m\\
\hline
PL at Reference Distance $\text{PL}_0$ &
$-30$ dB\\
\hline
PLE of Source-to-IRS Link $\gamma_{SI}$ & {$2.5$}\\
\hline
PLE of IRS-to-Destination Link $\gamma_{ID}$ & $2.5$\\
\hline
PLE of Source-to-Destination Link $\gamma_{SD}$ & $3.5$\\
\hline
Distance Between Source and IRS $d_{SI}$ & $50$ m \\
\hline
Vertical Distance Between Destination and Source-to-IRS Link $d_{v}$ & $2$ m \\
\hline
Horizontal Distance Between Source and Destination $d_{SD,h}$ & $49$ m \\
\hline
\end{tabular}
\end{table*}

\begin{figure}[t]
    \begin{center}
      \epsfxsize=7.0in\includegraphics[scale=0.7]{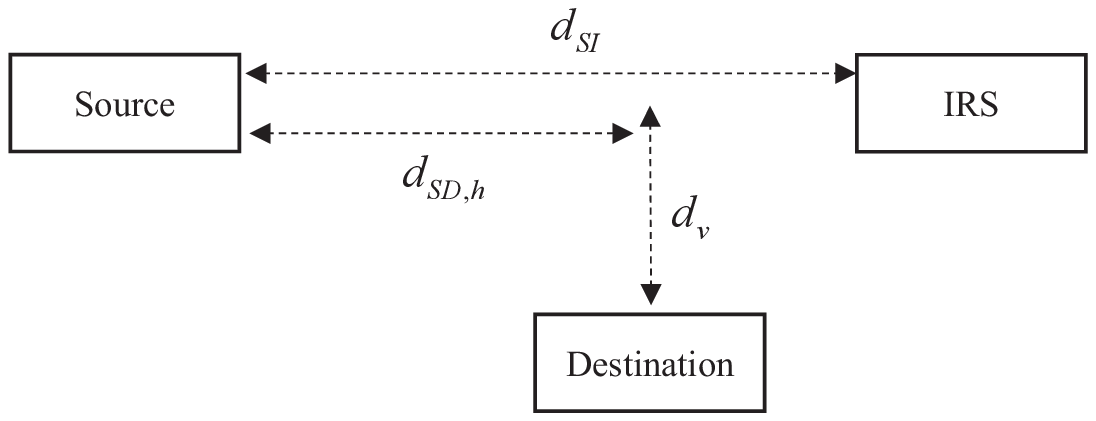}
      \caption{Simulation setup and the definitions of $d_{SI}$, $d_v$, and $d_{SD,h}$ \cite{Wu2018GLOBECCOM}.}\label{fig:distance}
    \end{center}
\end{figure}
We investigate both received SNR and symbol error rate (SER) performance for the above beamforming designs. For the SNR simulation, we obtain the results by averaging over 500 Rayleigh-faded channels, where we also adopt a theoretical SNR upper bound as a benchmark, which is calculated by solving problem \eqref{eq:SNRMaxProbwtheta5} and substituting the optimal objective value into \eqref{eq:psitheta}. {For the SER simulation, we generate 500 Rayleigh-faded channel realizations and  2,000 random quadrature phase shift keying (QPSK) symbols for each channel realization. Moreover, QPSK demodulation  at the destination is performed based on the equalized signal, i.e., $\frac{y}{(\mathbf{h}_{ID}^H\mathbf{\Theta}\mathbf{H}_{SI}+\mathbf{h}_{SD}^H)\mathbf w}$.}

During the simulation, the path loss (PL) of all the channels is modeled by
\begin{align}
\text{PL}=\left(\text{PL}_0-10\gamma\log_{10}\left(\frac{d}{d_0}\right)\right)\ \text{dB},
\end{align}
where $\text{PL}_0$ denotes the path loss at the reference distance $d_0$, $\gamma$ represents the path loss exponent (PLE), and $d$ stands for the distance between the transmitter and the receiver. The simulation parameters are set according to Table~\ref{table:simupara} unless otherwise specified, where the definitions of $d_{SI}$, $d_v$, and $d_{SD,h}$ are illustrated in Fig.~\ref{fig:distance}.
\begin{figure}
\centering
\subfigure[SNR Performance.]{      \epsfxsize=7.0in\includegraphics[scale=0.6]{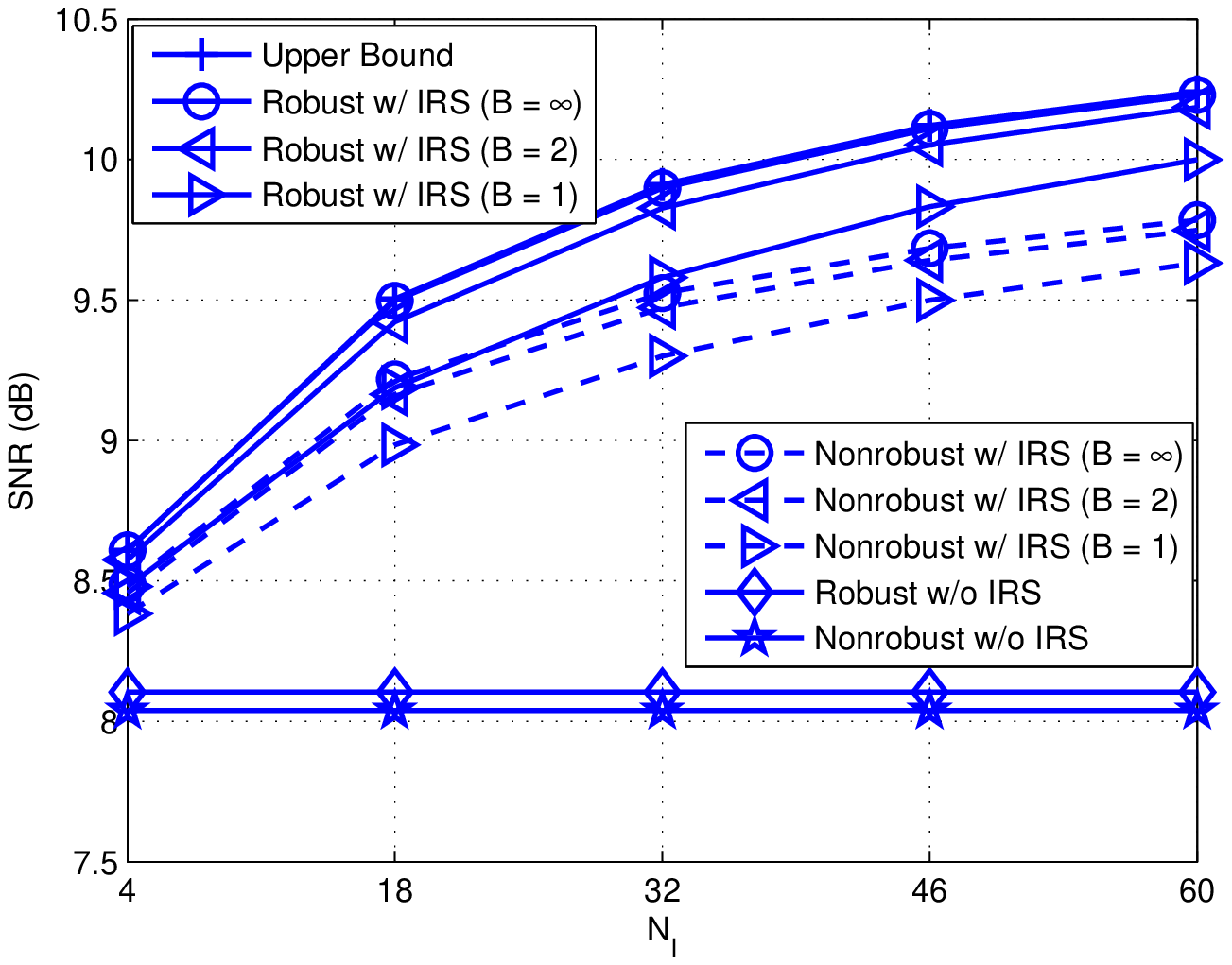}
     \label{fig:SNRN}}
    \subfigure[SER Performance.]{
      \epsfxsize=7.0in\includegraphics[scale=0.6]{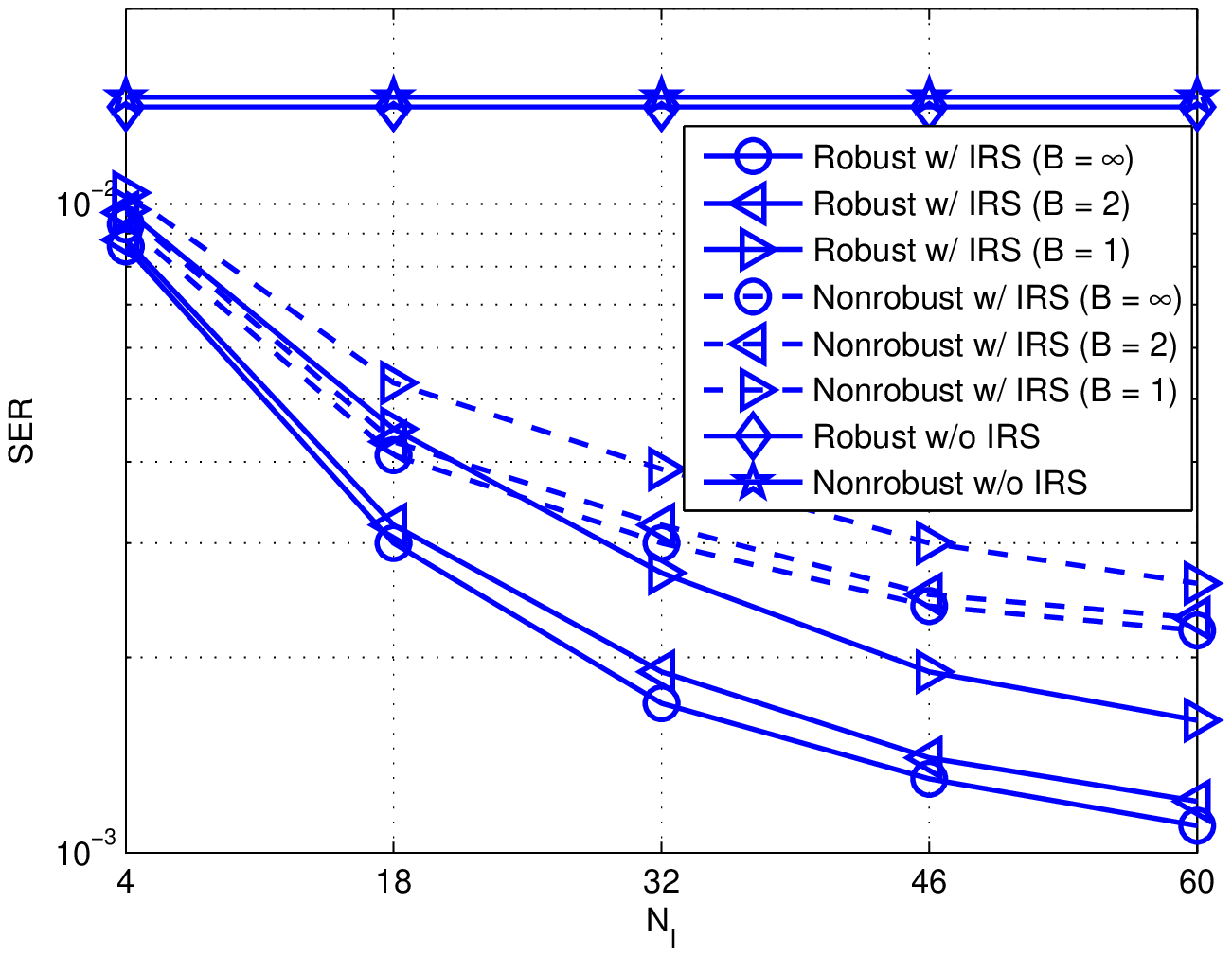}
      \label{fig:SERN}}
     \caption{SNR and SER versus the number of reflecting elements $N_I$ for different designs.}
\end{figure}
\begin{figure}
\centering
\subfigure[SNR Performance.]{      \epsfxsize=7.0in\includegraphics[scale=0.6]{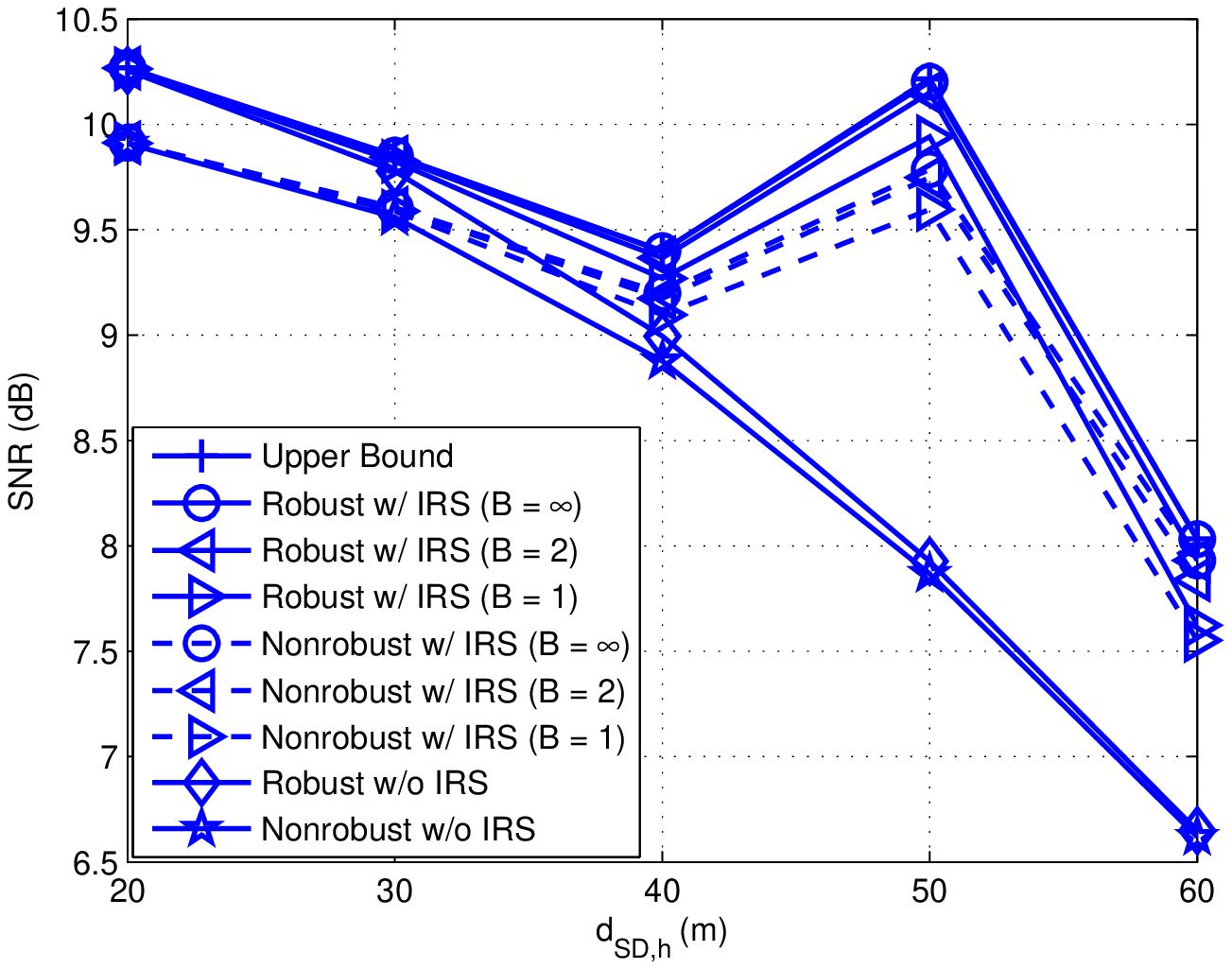}
     \label{fig:SNRD}}
    \subfigure[SER Performance.]{
      \epsfxsize=7.0in\includegraphics[scale=0.6]{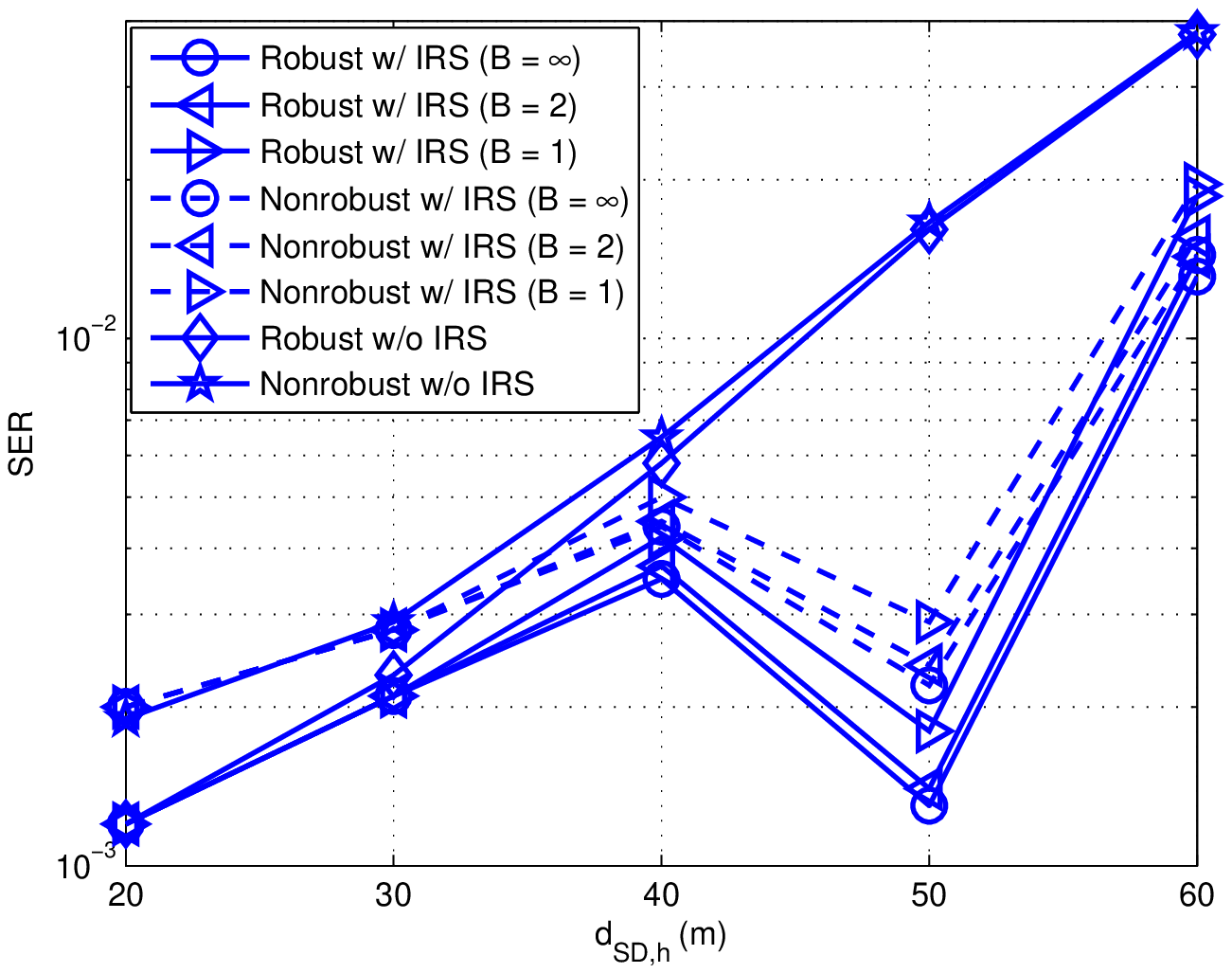}
      \label{fig:SERD}}
     \caption{SNR and SER versus the horizontal distance between source and destination $d_{SD,h}$  for different designs.}
\end{figure}


\subsection{{Simulation Results}}
Fig.~\ref{fig:SNRN} and Fig.~\ref{fig:SERN}  show the SNR and the SER of various beamforming designs versus $N_I$, respectively.  We can find from both figures that when $N_I$ increases, the performance of both robust and nonrobust designs for IRS-aided systems is improved due to the increasing beamforming gains brought by the IRS.  {Also, the proposed ``Robust w/ IRS'' outperforms the conventional ``Nonrobust w/ IRS'' by evident gains under both continuous and discrete phase shift cases. In particular, the gain is enlarged with the increase of $N_I$. In contrast, the performance advantage of ``Robust w/o IRS'' compared with ``Nonrobust w/o IRS'' is minor. This is due to the fact that both the IRS reflect beamforming and the source transmit beamforming designs are affected by the transceiver hardware impairments for the IRS-assisted communications.}
It is also worthwhile pointing out that ``Robust w/ IRS'' achieves almost the same received SNR as the theoretical upper bound in Fig.~\ref{fig:SNRN}, which validates the effectiveness of the proposed design. Moreover, compared to the ideal continuous phase shift case, the discrete phase shifts can lead to performance degradation especially for large $N_I$. However, the performance loss becomes minor when we adopt 2 bits to represent the IRS phase shifts. {This is because the obtained continuous solution is near optimal and it is possible to find an effective discrete solution by mapping the obtained continuous solution to the discrete phase set. Note that similar observations have been made in prior works such as \cite{Guo2019Arxiv}.}

\begin{figure}
\centering
\subfigure[SNR Performance.]{      \epsfxsize=7.0in\includegraphics[scale=0.6]{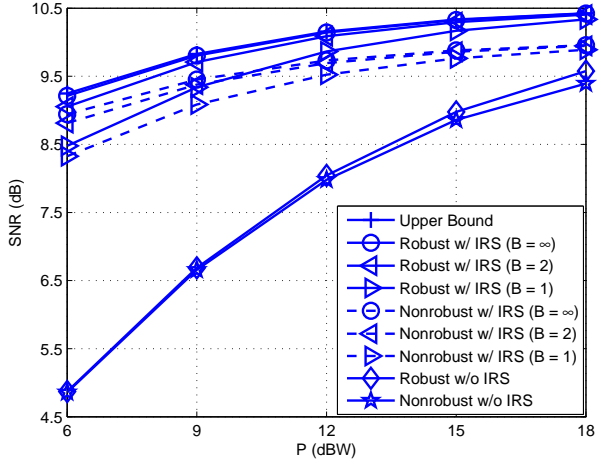}
     \label{fig:SNRP}}
    \subfigure[SER Performance.]{
      \epsfxsize=7.0in\includegraphics[scale=0.6]{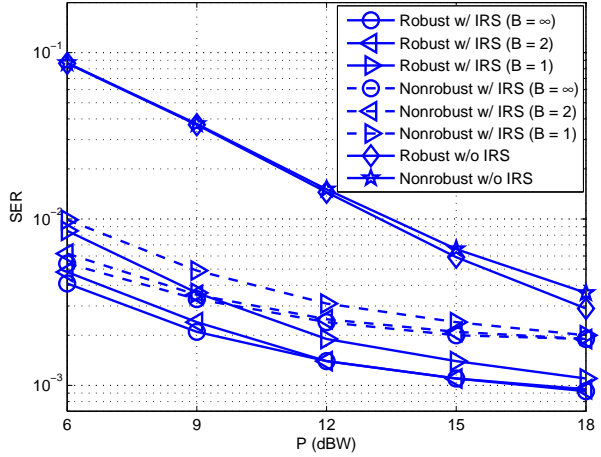}
      \label{fig:SERP}}
     \caption{SNR and SER versus source transmit power $P$  for different designs.}
\end{figure}
We compare the SNR and the SER of various schemes versus $d_{SD,h}$ in Fig.~\ref{fig:SNRD} and Fig.~\ref{fig:SERD}, respectively. As can be found from both figures, the performance of the conventional system without IRS degrades rapidly with the increase of $d_{SD,h}$ due to the increase of the associated path loss. In contrast, the performance of the IRS-assisted transmission first degrades, then improves, and finally degrades again when $d_{SD,h}$ becomes larger. It can be seen that the best performance is achieved when the destination is closest to the IRS, i.e., $d_{SD,h}=d_{SI}=50\ \text{m}$. This is because the introduction of an IRS helps to enhance the received signal strength via signal reflection when the destination gets close to the IRS, thus improving the system performance. Note that similar phenomenon has also been observed in \cite{Wu2018GLOBECCOM}.  We can also find from both figures that the proposed ``Robust w/ IRS'' outperforms ``Nonrobust w/ IRS'' under different values of $d_{SD,h}$, which again validates the necessity of incorporating the hardware impairments into the beamforming optimization.

\begin{figure}
\centering
\subfigure[SNR Performance.]{      \epsfxsize=7.0in\includegraphics[scale=0.6]{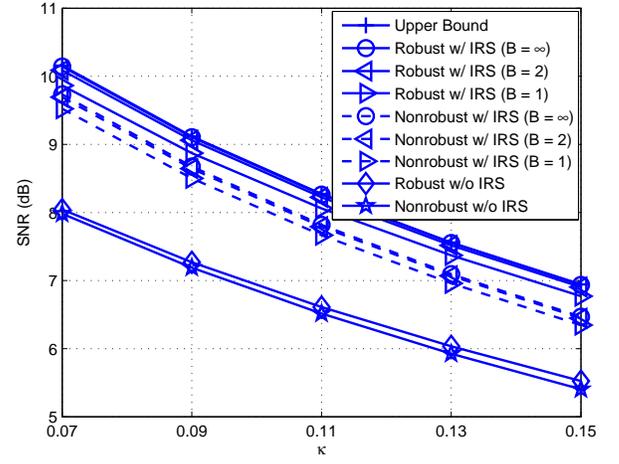}
     \label{fig:SNRK}}
    \subfigure[SER Performance.]{
      \epsfxsize=7.0in\includegraphics[scale=0.6]{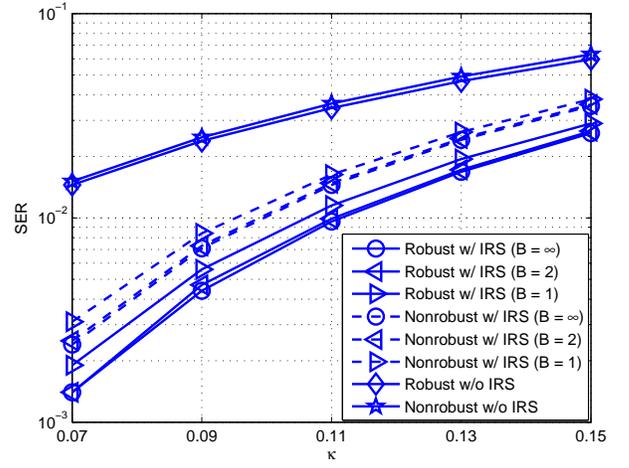}
      \label{fig:SERK}}
     \caption{SNR and SER versus normalized distortion power $\kappa$  for different designs.}
\end{figure}
The SNR and the SER versus $P$ are shown in Fig.~\ref{fig:SNRP} and Fig.~\ref{fig:SERP}, respectively. We can find that the performance of all the methods is enhanced as $P$ increases. Similarly to previous observations, the performance improvement brought by the schemes taking into account the hardware impairments is significant, particularly for the IRS-assisted transmission. Moreover, there exist error floors for the IRS-aided communications when $P$ is sufficiently large in Fig.~\ref{fig:SERP}, which is due to the fact that the power of the distortions caused by hardware impairments also grows with the increase of $P$, as predicted in \eqref{eq:SNR}. Nevertheless, the proposed optimization framework can effectively lower the error floors for a given transmit power $P$ compared with the nonrobust scheme.

We show the SNR and the SER comparisons versus $\kappa$ in Fig.~\ref{fig:SNRK} and Fig.~\ref{fig:SERK}, respectively. It can be seen that the performance of various designs deteriorates with the increase of $\kappa$ since the negative impact of hardware impairments becomes more prominent. It can also been found that large performance gains can be achieved due to the use of the IRS for relatively small $\kappa$. Yet, the gains become smaller for large $\kappa$ since the hardware impairments severely degrade the performance of both IRS-aided and conventional communications. In addition, as expected, by taking into account the hardware impairments, the proposed ``Robust w/ IRS'' outperforms ``Nonrobust w/ IRS'' under different $\kappa$'s.


\begin{figure*}[ht]
\normalsize
\newcounter{mytempeqncnt9}
\setcounter{mytempeqncnt9}{\value{equation}}
\setcounter{equation}{23}
\begin{align}\label{eq:woptinv}
& \left(\kappa_D\mathbf \Upsilon(\mathbf{\Theta})+ (1+\kappa_D)\kappa_S{\widetilde {\text{diag}}}\{\mathbf \Upsilon(\mathbf{\Theta})\}+\frac{(1+\kappa_D)\sigma_{n}^2}{\tilde P}\mathbf I_{N_S}\right)^{-1}\nonumber \\
= & \mathbf {\tilde \Upsilon}^{-1}(\mathbf{\Theta})-  \frac{\mathbf {\tilde \Upsilon}^{-1}(\mathbf{\Theta})(\mathbf{H}_{SI}^H\mathbf{\Theta}^H\mathbf{h}_{ID}+
\mathbf{h}_{SD})(\mathbf{h}_{ID}^H\mathbf{\Theta}\mathbf{H}_{SI}+
\mathbf{h}_{SD}^H)\mathbf {\tilde \Upsilon}^{-1}(\mathbf{\Theta})}{(\mathbf{h}_{ID}^H\mathbf{\Theta}\mathbf{H}_{SI}+
\mathbf{h}_{SD}^H)\mathbf {\tilde \Upsilon}^{-1}(\mathbf{\Theta})(\mathbf{H}_{SI}^H\mathbf{\Theta}^H\mathbf{h}_{ID}+
\mathbf{h}_{SD})+(1/\kappa_D)}
\end{align}
\setcounter{equation}{\value{mytempeqncnt9}}
\vspace*{4pt}
\hrulefill
\end{figure*}
\begin{figure*}[ht]
\normalsize
\newcounter{mytempeqncnt10}
\setcounter{mytempeqncnt10}{\value{equation}}
\setcounter{equation}{24}
\begin{align}\label{eq:woptinv1}
& \left(\kappa_D\mathbf \Upsilon(\mathbf{\Theta})+ \mathbf {\tilde \Upsilon}(\mathbf{\Theta})\right)^{-1}(\mathbf{H}_{SI}^H\mathbf{\Theta}^H\mathbf{h}_{ID}+
\mathbf{h}_{SD})\nonumber \\
=& \mathbf {\tilde \Upsilon}^{-1}(\mathbf{\Theta})(\mathbf{H}_{SI}^H\mathbf{\Theta}^H\mathbf{h}_{ID}+
\mathbf{h}_{SD}) -\frac{\mathbf {\tilde \Upsilon}^{-1}(\mathbf{\Theta})(\mathbf{H}_{SI}^H\mathbf{\Theta}^H\mathbf{h}_{ID}+
\mathbf{h}_{SD})(\mathbf{h}_{ID}^H\mathbf{\Theta}\mathbf{H}_{SI}+
\mathbf{h}_{SD}^H)\mathbf {\tilde \Upsilon}^{-1}(\mathbf{\Theta})(\mathbf{H}_{SI}^H\mathbf{\Theta}^H\mathbf{h}_{ID}+
\mathbf{h}_{SD})}{(\mathbf{h}_{ID}^H\mathbf{\Theta}\mathbf{H}_{SI}+
\mathbf{h}_{SD}^H)\mathbf {\tilde \Upsilon}^{-1}(\mathbf{\Theta})(\mathbf{H}_{SI}^H\mathbf{\Theta}^H\mathbf{h}_{ID}+
\mathbf{h}_{SD})+(1/\kappa_D)} \nonumber \\
= & \left(1- \frac{(\mathbf{h}_{ID}^H\mathbf{\Theta}\mathbf{H}_{SI}+
\mathbf{h}_{SD}^H)\mathbf {\tilde \Upsilon}^{-1}(\mathbf{\Theta})(\mathbf{H}_{SI}^H\mathbf{\Theta}^H\mathbf{h}_{ID}+
\mathbf{h}_{SD})}{(\mathbf{h}_{ID}^H\mathbf{\Theta}\mathbf{H}_{SI}+
\mathbf{h}_{SD}^H)\mathbf {\tilde \Upsilon}^{-1}(\mathbf{\Theta})(\mathbf{H}_{SI}^H\mathbf{\Theta}^H\mathbf{h}_{ID}+
\mathbf{h}_{SD})+(1/\kappa_D)}\right)\mathbf {\tilde \Upsilon}^{-1}(\mathbf{\Theta}) (\mathbf{H}_{SI}^H\mathbf{\Theta}^H\mathbf{h}_{ID}+
\mathbf{h}_{SD})\nonumber \\
= & \frac{1}{\kappa_D(\mathbf{h}_{ID}^H\mathbf{\Theta}\mathbf{H}_{SI}+
\mathbf{h}_{SD}^H)\mathbf {\tilde \Upsilon}^{-1}(\mathbf{\Theta})(\mathbf{H}_{SI}^H\mathbf{\Theta}^H\mathbf{h}_{ID}+
\mathbf{h}_{SD})+1}\mathbf {\tilde \Upsilon}^{-1}(\mathbf{\Theta})(\mathbf{H}_{SI}^H\mathbf{\Theta}^H\mathbf{h}_{ID}+
\mathbf{h}_{SD})
\end{align}
\setcounter{equation}{\value{mytempeqncnt10}}
\vspace*{4pt}
\hrulefill
\end{figure*}
\begin{figure*}[ht]
\normalsize
\newcounter{mytempeqncnt11}
\setcounter{mytempeqncnt11}{\value{equation}}
\setcounter{equation}{25}
\begin{align}\label{eq:SNRMaxProbwtheta1}
\mathop{\text{maximize}}\limits_{\boldsymbol \theta}  \quad &  (\boldsymbol \theta^H \text{diag}\{\mathbf{h}_{ID}^H\}\mathbf{H}_{SI}+
\mathbf{h}_{SD}^H)\left((1+\kappa_D)\kappa_S{\widetilde {\text{diag}}}\{\mathbf \Upsilon(\boldsymbol \theta)\} +\frac{(1+\kappa_D)\sigma_{n}^2}{\tilde P}\mathbf I_{N_S}\right)^{-1} (\mathbf{H}_{SI}^H\text{diag}\{\mathbf{h}_{ID}\}\boldsymbol \theta+
\mathbf{h}_{SD})\nonumber\\
\text{subject to}\quad & |\theta_i|=1,\ i=1,\cdots,N_I
\end{align}
\setcounter{equation}{\value{mytempeqncnt11}}
\vspace*{4pt}
\hrulefill
\end{figure*}
\begin{figure*}[ht]
\normalsize
\newcounter{mytempeqncnt12}
\setcounter{mytempeqncnt12}{\value{equation}}
\setcounter{equation}{26}
\begin{align}\label{eq:SNRMaxProbwtheta2}
\mathop{\text{maximize}}\limits_{\boldsymbol \theta,t}  \quad &  (\boldsymbol \theta^H \text{diag}\{\mathbf{h}_{ID}^H\}\mathbf{H}_{SI}+
t^*\mathbf{h}_{SD}^H)\left((1+\kappa_D)\kappa_S{\widetilde {\text{diag}}}\{\mathbf \Upsilon(\boldsymbol \theta,t) \}+\frac{(1+\kappa_D)\sigma_{n}^2}{\tilde P}\mathbf I_{N_S}\right)^{-1}  (\mathbf{H}_{SI}^H\text{diag}\{\mathbf{h}_{ID}\}\boldsymbol \theta+t\mathbf{h}_{SD})\nonumber\\
\text{subject to}\quad & |\theta_i|=1,\ i=1,\cdots,N_I,\nonumber \\
& |t|=1
\end{align}
\setcounter{equation}{\value{mytempeqncnt12}}
\vspace*{4pt}
\hrulefill
\end{figure*}
\subsection{Complexity Analysis}\label{sec:comp}
Now we compare the computational complexity of all the above considered beamforming schemes as shown in Table~\ref{table:comp}, $N_{\text{iter}}$ is the number of iterations required by ``Robust w/ IRS'', and $N_{\text{iter}}^{'}$ is the number of iterations needed by ``Nonrobust w/ IRS''. {It can be found that all the methods have polynomial-time computational complexity.}  The major computation of ``Robust w/ IRS'' in each iteration lies in the calculation of the vector $ (\mathbf \Psi^H \mathbf \Xi_0^{-1}\mathbf \Psi -(1+\kappa_D)\kappa_S(\mathbf \Omega-\lambda_\text{max}(\mathbf \Omega) \mathbf I_{N_I+1}))\boldsymbol {\tilde \theta}^{(k-1)}$ whose complexity is $\mathcal{O}(N_SN_I+N_I^2)$. Note that $\lambda_{\text{max}}(\mathbf \Omega)$ is computed via the  efficient power iteration method \cite{Wilkinson1965Book}. ``Nonrobust w/ IRS'' mainly needs to calculate the vector $\mathbf \Psi^H\mathbf \Psi \boldsymbol \theta^{(k-1)}$ in each iteration, whose complexity is $\mathcal{O}(N_SN_I)$.

We compare the average number of iterations required by ``Robust w/ IRS'' and ``Nonrobust w/ IRS'' in Table~\ref{table:iteration} for different values of $N_I$. It can be seen that the algorithm convergence becomes slower when $N_I$ increases. Moreover, the use of the acceleration scheme \cite[Section V-B]{Song2016TSP} results in faster convergence especially for ``Robust w/ IRS''. ``Robust w/ IRS'' needs more iterations to converge than ``Nonrobust w/ IRS'' while the gap becomes smaller after applying the acceleration method.

Based on the above simulation results and complexity analysis, we can conclude that the proposed robust design incorporating hardware impairments can strike a good balance of the tradeoff between performance and computational complexity.

\begin{figure*}[ht]
\normalsize
\newcounter{mytempeqncnt13}
\setcounter{mytempeqncnt13}{\value{equation}}
\setcounter{equation}{27}
\begin{align}\label{eq:SNRMaxProbwtheta3}
\mathop{\text{maximize}}\limits_{\boldsymbol {\tilde \theta}}  \quad &  \boldsymbol {\tilde \theta}^H\mathbf \Psi^H\left((1+\kappa_D)\kappa_S{\widetilde {\text{diag}}}\{\mathbf \Psi \boldsymbol {\tilde \theta}\boldsymbol {\tilde \theta}^H\mathbf \Psi^H\}+\frac{(1+\kappa_D)\sigma_{n}^2}{\tilde P}\mathbf I_{N_S}\right)^{-1} \mathbf \Psi \boldsymbol {\tilde \theta}\nonumber\\
\text{subject to}\quad & |\tilde \theta_i|=1,\ i=1,\cdots,N_I+1
\end{align}
\setcounter{equation}{\value{mytempeqncnt13}}
\vspace*{4pt}
\hrulefill
\end{figure*}
\begin{figure*}[ht]
\normalsize
\newcounter{mytempeqncnt14}
\setcounter{mytempeqncnt14}{\value{equation}}
\setcounter{equation}{28}
\begin{align}\label{eq:objSDR}
& [(\mathbf \Psi \boldsymbol {\tilde \theta})_1^*,\cdots,(\mathbf \Psi \boldsymbol {\tilde \theta})_{N_S}^*]  \text{diag}\left\{\left[\left((1+\kappa_D)\kappa_S(\mathbf \Psi \boldsymbol {\tilde \theta}\boldsymbol {\tilde \theta}^H\mathbf \Psi^H)_{1,1}+\frac{(1+\kappa_D)\sigma_{n}^2}{\tilde P}\right)^{-1},\cdots, \right.\right.\nonumber \\ &\left. \left. \left((1+\kappa_D)\kappa_S(\mathbf \Psi \boldsymbol {\tilde \theta}\boldsymbol {\tilde \theta}^H\mathbf \Psi^H)_{N_S,N_S}+\frac{(1+\kappa_D)\sigma_{n}^2}{\tilde P}\right)^{-1}\right]\right\} [(\mathbf \Psi \boldsymbol {\tilde \theta})_1,\cdots,(\mathbf \Psi \boldsymbol {\tilde \theta})_{N_S}]^T \nonumber \\ & = \sum_{m=1}^{N_S} \frac{|(\mathbf \Psi \boldsymbol {\tilde \theta})_m|^2}{(1+\kappa_D)\kappa_S(\mathbf \Psi \boldsymbol {\tilde \theta}\boldsymbol {\tilde \theta}^H\mathbf \Psi^H)_{m,m}+\frac{(1+\kappa_D)\sigma_{n}^2}{\tilde P}}
\end{align}
\setcounter{equation}{\value{mytempeqncnt14}}
\vspace*{4pt}
\hrulefill
\end{figure*}

\begin{table*}[t]
\renewcommand{\arraystretch}{1.3}
\caption{Beamforming Designs Complexity}
\label{table:comp}
\centering
\begin{tabular}{|c|c|}
\hline
Beamforming Design Method & Computational Complexity\\
\hline
Robust w/ IRS & {$\mathcal{O}(N_{\text{iter}}(N_SN_I+N_I^2))$}\\
\hline
Nonrobust w/ IRS & {$\mathcal{O}(N_{\text{iter}}^{'}N_SN_I)$}\\
\hline
Robust w/o IRS & $\mathcal{O}(N_S)$\\
\hline
Nonrobust w/o IRS & $\mathcal{O}(N_S)$\\
\hline
\end{tabular}
\end{table*}

\begin{table*}[t]
\renewcommand{\arraystretch}{1.3}
\caption{Comparison of Average Number of Iterations (Convergence Accuracy $\epsilon=10^{-5}$)}
\label{table:iteration}
\centering
\begin{tabular}{|c|c|c|c|c|c|}
\hline
\multirow{2}*{Beamforming Design Method} & \multicolumn{5}{c|}{Average Number of Iterations} \\
\cline{2-6}
& $N_I=4$ & $N_I=18$ & $N_I=32$ & $N_I=46$ & $N_I=60$ \\
\hline
Robust w/o Acceleration & $161.69 $  & $280.61 $ &   $ 411.24$  &  $492.84$ &   $518.17$  \\
\hline
Robust w/ Acceleration & $8.31$ & $19.92 $ & $ 27.1 $ &  $ 33.44 $  & $36.57$\\
\hline
Nonrobust w/o Acceleration & $5.13  $ & $ 9.82 $ & $ 14.63 $ &  $  15.65  $  & $  17.83$       \\
\hline
Nonrobust w/ Acceleration & $2.97 $ & $ 4.31 $ & $5.73 $ &  $6.2$  & $ 6.52$  \\
\hline
\end{tabular}
\end{table*}

\section{Conclusions}
We investigated the beamforming optimization for IRS-aided multi-antenna communications in the presence of transceiver hardware impairments. By incorporating signal distortions caused by the hardware impairments, we jointly optimized the source transmit beamforming and IRS reflect beamforming to maximize the received SNR at the destination. The optimal solution to the source transmit beamforming was obtained in a closed form, which turns out to be weighted MF beamforming. An MM-based algorithm was developed to optimize the IRS reflect beamforming, which approaches the theoretical upper bound with low computational complexity. The performance advantage of the proposed beamforming design was also validated via extensive simulations and the importance of considering hardware impairments in the beamforming design was unveiled.

\begin{appendices}
\section{Proof of {Proposition}~\ref{prop:wopt}}\label{app:wopt}
We first apply the matrix inversion lemma \cite{Horn1985} and rewrite the matrix inversion term in \eqref{eq:wopt} by \eqref{eq:woptinv},
where $\mathbf {\tilde \Upsilon}(\mathbf{\Theta}) \triangleq (1+\kappa_D)\kappa_S{\widetilde {\text{diag}}}\{\mathbf \Upsilon(\mathbf{\Theta})\}+\frac{(1+\kappa_D)\sigma_{n}^2}{\tilde P}\mathbf I_{N_S}$. Then, it follows that \eqref{eq:woptinv1} holds.
Substituting \eqref{eq:woptinv1} into \eqref{eq:wopt}, we obtain \eqref{eq:wopt1}.

\section{Proof of Theorem~\ref{theo:upper}}\label{app:theoupper}
We first rewrite problem \eqref{eq:SNRMaxProbwtheta0} by \eqref{eq:SNRMaxProbwtheta1},
where $\boldsymbol \theta\triangleq [\theta_1^*,\cdots,\theta_{N_I}^*]^T$ and {$\mathbf \Upsilon(\boldsymbol \theta) \triangleq(\mathbf{H}_{SI}^H\text{diag}\{\mathbf{h}_{ID}\}\boldsymbol \theta+
\mathbf{h}_{SD})(\boldsymbol \theta^H \text{diag}\{\mathbf{h}_{ID}^H\}\mathbf{H}_{SI}+
\mathbf{h}_{SD}^H)$}.
Then, we introduce a complex slack variable {$t$} satisfying $|t|=1$ and recast the above problem as \eqref{eq:SNRMaxProbwtheta2},
where {$\mathbf \Upsilon(\boldsymbol \theta,t) \triangleq(\mathbf{H}_{SI}^H\text{diag}\{\mathbf{h}_{ID}\}\boldsymbol \theta+
t\mathbf{h}_{SD})(\boldsymbol \theta^H \text{diag}\{\mathbf{h}_{ID}^H\}\mathbf{H}_{SI}+
t^*\mathbf{h}_{SD}^H)$}. {Since $|t|=1$, it can be readily confirmed that, if $(\boldsymbol{\theta}^\star,t^\star)$ is optimal to problem \eqref{eq:SNRMaxProbwtheta2}, ${\boldsymbol{\theta}^\star}/{t^\star}$ will also be optimal to problem \eqref{eq:SNRMaxProbwtheta1} and vice versa. Therefore, problems \eqref{eq:SNRMaxProbwtheta1} and \eqref{eq:SNRMaxProbwtheta2} are equivalent.}

Define $\boldsymbol {\tilde \theta}\triangleq[\boldsymbol {\theta}^T\ t]^T$ and $\mathbf \Psi\triangleq[\mathbf{H}_{SI}^H\text{diag}\{\mathbf{h}_{ID}\}\ \mathbf{h}_{SD}]$. Then, we rewrite problem \eqref{eq:SNRMaxProbwtheta2} by \eqref{eq:SNRMaxProbwtheta3} on the top of next page,
where $\tilde \theta_i$ is the $i$-th entry of $\boldsymbol {\tilde \theta}$.
After some matrix manipulations, we express the objective function of problem \eqref{eq:SNRMaxProbwtheta3} by \eqref{eq:objSDR} on the top of next page.
Since $|(\mathbf \Psi \boldsymbol {\tilde \theta})_m|^2=(\mathbf \Psi \boldsymbol {\tilde \theta}\boldsymbol {\tilde \theta}^H\mathbf \Psi^H)_{m,m}$, problem \eqref{eq:SNRMaxProbwtheta3} now becomes
\setcounter{equation}{29}
\begin{align}\label{eq:SNRMaxProbwtheta4}
\mathop{\text{maximize}}\limits_{\boldsymbol {\tilde \theta}}  \quad &  \sum_{m=1}^{N_S} \frac{(\mathbf \Psi \boldsymbol {\tilde \theta}\boldsymbol {\tilde \theta}^H\mathbf \Psi^H)_{m,m}}{(1+\kappa_D)\kappa_S(\mathbf \Psi \boldsymbol {\tilde \theta}\boldsymbol {\tilde \theta}^H\mathbf \Psi^H)_{m,m}+\frac{(1+\kappa_D)\sigma_{n}^2}{\tilde P}}\nonumber\\
\text{subject to}\quad & |\tilde \theta_i|=1,\ i=1,\cdots,N_I+1.
\end{align}
By defining $\boldsymbol {\tilde \Theta}\triangleq\boldsymbol {\tilde \theta}\boldsymbol {\tilde \theta}^H$ and relaxing the rank-one constraint with respect to $\boldsymbol {\tilde \Theta}$, we obtain problem \eqref{eq:SNRMaxProbwtheta5}, i.e.,
 \begin{align}
\mathop{\text{maximize}}\limits_{\boldsymbol {\tilde \Theta} \succeq \mathbf 0}  \quad &  \sum_{m=1}^{N_S} \frac{(\mathbf \Psi \boldsymbol {\tilde \Theta}\mathbf \Psi^H)_{m,m}}{(1+\kappa_D)\kappa_S(\mathbf \Psi \boldsymbol {\tilde \Theta}\mathbf \Psi^H)_{m,m}+\frac{(1+\kappa_D)\sigma_{n}^2}{\tilde P}}\nonumber\\
\text{subject to}\quad & (\boldsymbol {\tilde \Theta})_{i,i}=1,\ i=1,\cdots,N_I+1.
\end{align}
Clearly, the optimal objective value of this problem must be an upper bound to that of problem \eqref{eq:SNRMaxProbwtheta0} due to the relaxation of the rank-one constraint.

To prove that the above problem is convex, we only need to show that the objective function is concave with respect to $\mathbf {\tilde \Theta}$. We express the objective function by
\begin{align}
& \sum_{m=1}^{N_S} \frac{(\mathbf \Psi \boldsymbol {\tilde \Theta}\mathbf \Psi^H)_{m,m}}{(1+\kappa_D)\kappa_S(\mathbf \Psi \boldsymbol {\tilde \Theta}\mathbf \Psi^H)_{m,m}+\frac{(1+\kappa_D)\sigma_{n}^2}{\tilde P}}\!=\! \frac{1}{(1+\kappa_D)\kappa_S} \nonumber \\ & \times \sum_{m=1}^{N_S} \left(1\!-\!\frac{\frac{(1+\kappa_D)\sigma_{n}^2}{\tilde P}}{(1\!+\!\kappa_D)\kappa_S(\mathbf \Psi \boldsymbol {\tilde \Theta}\mathbf \Psi^H)_{m,m}\!+\!\frac{(1+\kappa_D)\sigma_{n}^2}{\tilde P}}\right).
\end{align}
Define $q_m\triangleq(\mathbf \Psi \boldsymbol {\tilde \Theta}\mathbf \Psi^H)_{m,m}$ and $u(q_m)\triangleq-\frac{1}{(1+\kappa_D)\kappa_S q_m+\frac{(1+\kappa_D)\sigma_{n}^2}{\tilde P}}$. Then, it can be readily verified that $u(q_m)$ is concave. Moreover, $q_m$ is an affine function of $\boldsymbol {\tilde \Theta}$. Therefore, the objective function is also concave according to \cite[Section 3.2.2]{BoydConvex}.

\begin{figure*}[ht]
\normalsize
\newcounter{mytempeqncnt15}
\setcounter{mytempeqncnt15}{\value{equation}}
\setcounter{equation}{33}
\begin{align}\label{eq:fbnd}
\hat f(\boldsymbol {\tilde \theta} ,\mathbf \Xi,\boldsymbol {\tilde \theta}_0 ,\mathbf \Xi_0)& \overset{(a)}{=} \boldsymbol {\tilde \theta}_0^H\mathbf \Psi^H \mathbf \Xi_0^{-1}\mathbf \Psi \boldsymbol {\tilde \theta}_0+2\Re\{\boldsymbol {\tilde \theta}_0^H\mathbf \Psi^H \mathbf \Xi_0^{-1}\mathbf \Psi (\boldsymbol {\tilde \theta}-\boldsymbol {\tilde \theta}_0)\}+\text{tr}(-\mathbf \Xi_0^{-1}(\mathbf \Psi  \boldsymbol {\tilde \theta}_0\boldsymbol {\tilde \theta}_0^H\mathbf \Psi^H)^T\mathbf \Xi_0^{-1}\nonumber \\
& \ \ \ \times ((1+\kappa_D)\kappa_S{\widetilde {\text{diag}}}\{\mathbf \Psi \boldsymbol {\tilde \theta}\boldsymbol {\tilde \theta}^H\mathbf \Psi^H\}-(1+\kappa_D)\kappa_S{\widetilde {\text{diag}}}\{\mathbf \Psi \boldsymbol {\tilde \theta}_0\boldsymbol {\tilde \theta}_0^H\mathbf \Psi^H\}))\nonumber \\
& = 2\Re\{\boldsymbol {\tilde \theta}_0^H\mathbf \Psi^H \mathbf \Xi_0^{-1}\mathbf \Psi \boldsymbol {\tilde \theta}\} \!-\!(1\!+\!\kappa_D)\kappa_S \text{tr}(\mathbf \Xi_0^{-1}(\mathbf \Psi  \boldsymbol {\tilde \theta}_0\boldsymbol {\tilde \theta}_0^H\mathbf \Psi^H)^T\mathbf \Xi_0^{-1}{\widetilde {\text{diag}}}\{\mathbf \Psi \boldsymbol {\tilde \theta}\boldsymbol {\tilde \theta}^H\mathbf \Psi^H\} ) \nonumber \\
& \ \ \ -\boldsymbol {\tilde \theta}_0^H\mathbf \Psi^H \mathbf \Xi_0^{-1}\mathbf \Psi \boldsymbol {\tilde \theta}_0\!+\!(1+\kappa_D)\kappa_S \text{tr}(\mathbf \Xi_0^{-1}(\mathbf \Psi  \boldsymbol {\tilde \theta}_0\boldsymbol {\tilde \theta}_0^H\mathbf \Psi^H)^T\mathbf \Xi_0^{-1}{\widetilde {\text{diag}}}\{\mathbf \Psi \boldsymbol {\tilde \theta}_0\boldsymbol {\tilde \theta}_0^H\mathbf \Psi^H\} ) \nonumber \\
& \overset{(b)}{=} 2\Re\{\boldsymbol {\tilde \theta}_0^H\mathbf \Psi^H \mathbf \Xi_0^{-1}\mathbf \Psi \boldsymbol {\tilde \theta}\} -(1+\kappa_D)\kappa_S \boldsymbol {\tilde \theta}_0^H\mathbf \Psi^H\mathbf \Xi_0^{-1}{\widetilde {\text{diag}}}\{\mathbf \Psi \boldsymbol {\tilde \theta}\boldsymbol {\tilde \theta}^H\mathbf \Psi^H\}\mathbf \Xi_0^{-1}\mathbf \Psi  \boldsymbol {\tilde \theta}_0  \nonumber \\
& \ \ \ -\boldsymbol {\tilde \theta}_0^H\mathbf \Psi^H \mathbf \Xi_0^{-1}\mathbf \Psi \boldsymbol {\tilde \theta}_0+(1+\kappa_D)\kappa_S\boldsymbol {\tilde \theta}_0^H\mathbf \Psi^H\mathbf \Xi_0^{-1}{\widetilde {\text{diag}}}\{\mathbf \Psi \boldsymbol {\tilde \theta}_0\boldsymbol {\tilde \theta}_0^H\mathbf \Psi^H\}\mathbf \Xi_0^{-1}\mathbf \Psi  \boldsymbol {\tilde \theta}_0\nonumber \\
& \overset{(c)}{=} 2\Re\{\boldsymbol {\tilde \theta}_0^H\mathbf \Psi^H \mathbf \Xi_0^{-1}\mathbf \Psi \boldsymbol {\tilde \theta}\} -(1+\kappa_D)\kappa_S \boldsymbol {\tilde \theta}^H\mathbf \Psi^H {\widetilde {\text{diag}}}\{\mathbf \Xi_0^{-1}\mathbf \Psi  \boldsymbol {\tilde \theta}_0  \boldsymbol {\tilde \theta}_0^H\mathbf \Psi^H\mathbf \Xi_0^{-1}\} \mathbf \Psi \boldsymbol {\tilde \theta} \nonumber \\
& \ \ \ -\boldsymbol {\tilde \theta}_0^H\mathbf \Psi^H \mathbf \Xi_0^{-1}\mathbf \Psi \boldsymbol {\tilde \theta}_0+(1+\kappa_D)\kappa_S \boldsymbol {\tilde \theta}_0^H\mathbf \Psi^H {\widetilde {\text{diag}}}\{\mathbf \Xi_0^{-1}\mathbf \Psi  \boldsymbol {\tilde \theta}_0  \boldsymbol {\tilde \theta}_0^H\mathbf \Psi^H\mathbf \Xi_0^{-1}\} \mathbf \Psi \boldsymbol {\tilde \theta}_0 \nonumber \\
& \triangleq \hat f(\boldsymbol {\tilde \theta} ,\boldsymbol {\tilde \theta}_0 ,\mathbf \Xi_0)
\end{align}
\setcounter{equation}{\value{mytempeqncnt15}}
\vspace*{4pt}
\hrulefill
\end{figure*}
\begin{figure*}[ht]
\normalsize
\newcounter{mytempeqncnt16}
\setcounter{mytempeqncnt16}{\value{equation}}
\setcounter{equation}{35}
\begin{align}\label{eq:fbar}
\bar f(\boldsymbol {\tilde \theta}_0;\boldsymbol {\tilde \theta}_0)=&2\Re\{\boldsymbol {\tilde \theta}_0^H(\mathbf \Psi^H \mathbf \Xi_0^{-1}\mathbf \Psi -(1+\kappa_D)\kappa_S(\mathbf \Omega-\lambda_\text{max}(\mathbf \Omega)\mathbf I_{N_I+1})) \boldsymbol {\tilde \theta}_0\}-2(1+\kappa_D)\kappa_S(N_I+1)\lambda_\text{max}(\mathbf \Omega)\nonumber \\
& +\boldsymbol {\tilde \theta}_0^H(2(1+\kappa_D)\kappa_S\mathbf \Omega-\mathbf \Psi^H \mathbf \Xi_0^{-1}\mathbf \Psi)\boldsymbol {\tilde \theta}_0\nonumber \\
=&\boldsymbol {\tilde \theta}_0^H\mathbf \Psi^H \mathbf \Xi_0^{-1}\mathbf \Psi\boldsymbol {\tilde \theta}_0+2(1+\kappa_D)\kappa_S\|\boldsymbol {\tilde \theta}_0\|^2\lambda_\text{max}(\mathbf \Omega)-2(1+\kappa_D)\kappa_S(N_I+1)\lambda_\text{max}(\mathbf \Omega)\nonumber \\
\overset{(a)}{=}&\boldsymbol {\tilde \theta}_0^H\mathbf \Psi^H \mathbf \Xi_0^{-1}\mathbf \Psi\boldsymbol {\tilde \theta}_0\nonumber \\
\overset{(b)}{=}& f(\boldsymbol {\tilde \theta}_0)
\end{align}
\setcounter{equation}{\value{mytempeqncnt16}}
\vspace*{4pt}
\hrulefill
\end{figure*}
\section{Proof of Proposition~\ref{prop:MMbnd}}\label{app:MMbnd}
Define $\mathbf \Xi\triangleq(1+\kappa_D)\kappa_S{\widetilde {\text{diag}}}\{\mathbf \Psi \boldsymbol {\tilde \theta}\boldsymbol {\tilde \theta}^H\mathbf \Psi^H\}+\frac{(1+\kappa_D)\sigma_{n}^2}{\tilde P}\mathbf I_{N_S}$. Since the function $f(\mathbf x,\mathbf Y)=\mathbf x^H \mathbf Y^{-1} \mathbf x$ is jointly convex with $(\mathbf x,\mathbf Y)$ for $\mathbf Y \succ \mathbf 0$ \cite[Section 3.1.7]{BoydConvex} and $\mathbf \Psi \boldsymbol {\tilde \theta}$ is an affine transformation of $\boldsymbol {\tilde \theta}$, $f(\boldsymbol {\tilde \theta} ,\mathbf \Xi)=\boldsymbol {\tilde \theta}^H\mathbf \Psi^H \mathbf \Xi^{-1}\mathbf \Psi \boldsymbol {\tilde \theta} $ is also jointly convex with $(\boldsymbol {\tilde \theta},\mathbf \Xi)$ \cite[Section 3.2.2]{BoydConvex}. Recall that a convex function is lower bounded by its first-order Taylor expansion. Hence, by applying the complex-valued derivatives in \cite[Section 4.2]{Hjoungnes2011Book}, we  have
\begin{align}
f(\boldsymbol {\tilde \theta} ,\mathbf \Xi)
&  {\geq}  \boldsymbol {\tilde \theta}_0^H\mathbf \Psi^H \mathbf \Xi_0^{-1}\mathbf \Psi \boldsymbol {\tilde \theta}_0\!+\!2\Re\{\boldsymbol {\tilde \theta}_0^H\mathbf \Psi^H \mathbf \Xi_0^{-1}\mathbf \Psi (\boldsymbol {\tilde \theta}-\boldsymbol {\tilde \theta}_0)\}\nonumber \\ &+\!\text{tr}(-\mathbf \Xi_0^{-1}(\mathbf \Psi  \boldsymbol {\tilde \theta}_0\boldsymbol {\tilde \theta}_0^H\mathbf \Psi^H)^T\mathbf \Xi_0^{-1}(\mathbf \Xi-\mathbf \Xi_0))\nonumber \\
& \triangleq \hat f(\boldsymbol {\tilde \theta} ,\mathbf \Xi,\boldsymbol {\tilde \theta}_0 ,\mathbf \Xi_0),
\end{align}
where $\mathbf \Xi_0\triangleq(1+\kappa_D)\kappa_S{\widetilde {\text{diag}}}\{\mathbf \Psi \boldsymbol {\tilde \theta}_0\boldsymbol {\tilde \theta}_0^H\mathbf \Psi^H\}+\frac{(1+\kappa_D)\sigma_{n}^2}{\tilde P}\mathbf I_{N_S}$ and  $\boldsymbol {\tilde \theta}_0$ is a feasible point. Furthermore, we express $\hat f(\boldsymbol {\tilde \theta} ,\mathbf \Xi,\boldsymbol {\tilde \theta}_0 ,\mathbf \Xi_0)$ by \eqref{eq:fbnd},
where (a) follows from the definitions of $\mathbf \Xi$ and $\mathbf \Xi_0$, (b) holds due to $\text{tr}(\mathbf A\mathbf B)=\text{tr}(\mathbf B\mathbf A)$, and (c) follows from the equation $\mathbf a^H {\widetilde {\text{diag}}}\{\mathbf b\mathbf b^H\}\mathbf a=\mathbf b^H {\widetilde {\text{diag}}}\{\mathbf a\mathbf a^H\}\mathbf b$.

According to the inequality given by \cite[Eq. (26)]{Sun2017TSP}, the quadratic term in $\hat f(\boldsymbol {\tilde \theta} ,\boldsymbol {\tilde \theta}_0 ,\mathbf \Xi_0)$ is upper bounded by
\setcounter{equation}{34}
\begin{align}
& \boldsymbol {\tilde \theta}^H\mathbf \Psi^H {\widetilde {\text{diag}}}\{  \mathbf \Xi_0^{-1}\mathbf \Psi  \boldsymbol {\tilde \theta}_0  \boldsymbol {\tilde \theta}_0^H\mathbf \Psi^H\mathbf \Xi_0^{-1}\} \mathbf \Psi \boldsymbol {\tilde \theta}\nonumber \\
\overset{(a)}{\leq} & \lambda_\text{max}(\mathbf \Omega)\|\boldsymbol {\tilde \theta}\|^2+2\Re\{\boldsymbol {\tilde \theta}_0^H(\mathbf \Omega-\lambda_\text{max}(\mathbf \Omega)\mathbf I_{N_I+1})\boldsymbol {\tilde \theta}\}\nonumber \\&+\boldsymbol {\tilde \theta}_0^H(\lambda_\text{max}(\mathbf \Omega)\mathbf I_{N_I+1}-\mathbf \Omega)\boldsymbol {\tilde \theta}_0\nonumber \\
\overset{(b)}{=} & 2\Re\{\boldsymbol {\tilde \theta}_0^H(\mathbf \Omega-\lambda_\text{max}(\mathbf \Omega)\mathbf I_{N_I+1})\boldsymbol {\tilde \theta}\}+2(N_I+1)\lambda_\text{max}(\mathbf \Omega)\nonumber \\&-\boldsymbol {\tilde \theta}_0^H\mathbf \Omega\boldsymbol {\tilde \theta}_0,
\end{align}
where $\mathbf \Omega\triangleq\mathbf \Psi^H {\widetilde {\text{diag}}}\{ \mathbf \Xi_0^{-1}\mathbf \Psi  \boldsymbol {\tilde \theta}_0  \boldsymbol {\tilde \theta}_0^H\mathbf \Psi^H\mathbf \Xi_0^{-1}\} \mathbf \Psi$ in (a) and (b) holds because $|\tilde \theta_i|=|(\tilde \theta_0)_i|=1,\ i=1,\cdots,N_I+1$. By substituting the above inequality into \eqref{eq:fbnd}, we obtain \eqref{eq:fbndfinal}.

\begin{figure*}[ht]
\normalsize
\newcounter{mytempeqncnt17}
\setcounter{mytempeqncnt17}{\value{equation}}
\setcounter{equation}{39}
\begin{align}\label{eq:nablabarf}
\nabla \bar f(\boldsymbol {\tilde \theta};\boldsymbol {\tilde \theta}_0)=&2\boldsymbol {\tilde \theta}_0^H(\mathbf \Psi^H \mathbf \Xi_0^{-1}\mathbf \Psi -(1+\kappa_D)\kappa_S(\mathbf \Omega-\lambda_\text{max}(\mathbf \Omega)\mathbf I_{N_I+1}))-2(1+\kappa_D)\kappa_S\lambda_\text{max}(\mathbf \Omega)\boldsymbol {\tilde \theta}^H\nonumber \\
=& 2\boldsymbol {\tilde \theta}_0^H(\mathbf \Psi^H \mathbf \Xi_0^{-1}\mathbf \Psi -(1+\kappa_D)\kappa_S\mathbf \Omega)+2(1+\kappa_D)\kappa_S\lambda_\text{max}(\mathbf \Omega)(\boldsymbol {\tilde \theta}_0^H-\boldsymbol {\tilde \theta}^H)\nonumber \\
\overset{(a)}{=}& 2\boldsymbol {\tilde \theta}_0^H \mathbf \Psi^H (\mathbf \Xi_0^{-1}-(1+\kappa_D)\kappa_S{\widetilde {\text{diag}}}\{\mathbf \Xi_0^{-1}\mathbf \Psi  \boldsymbol {\tilde \theta}_0  \boldsymbol {\tilde \theta}_0^H\mathbf \Psi^H\mathbf \Xi_0^{-1}\}) \mathbf \Psi+2(1+\kappa_D)\kappa_S\lambda_\text{max}(\mathbf \Omega) (\boldsymbol {\tilde \theta}_0^H-\boldsymbol {\tilde \theta}^H)\nonumber \\ \overset{(b)}{=} & 2 \sum_{m=1}^{N_S} \boldsymbol {\tilde \theta}_0^H \mathbf \Psi^H  \mathbf e_m \mathbf e_m^T \mathbf \Psi \left(\frac{1}{(1+\kappa_D)\kappa_S\|\mathbf e_m^T\mathbf \Psi \boldsymbol {\tilde \theta}_0\|^2+\frac{(1+\kappa_D)\sigma_{n}^2}{\tilde P}}  -\frac{(1+\kappa_D)\kappa_S\|\mathbf e_m^T\mathbf \Psi \boldsymbol {\tilde \theta}_0\|^2}{\left((1+\kappa_D)\kappa_S\|\mathbf e_m^T\mathbf \Psi \boldsymbol {\tilde \theta}_0\|^2+\frac{(1+\kappa_D)\sigma_{n}^2}{\tilde P} \right)^2}\right)\nonumber \\ &+2(1+\kappa_D)\kappa_S\lambda_\text{max}(\mathbf \Omega) (\boldsymbol {\tilde \theta}_0^H-\boldsymbol {\tilde \theta}^H)\nonumber \\= & \frac{2(1+\kappa_D)\sigma_{n}^2}{\tilde P}\sum_{m=1}^{N_S} \frac{\boldsymbol {\tilde \theta}_0^H \mathbf \Psi^H  \mathbf e_m \mathbf e_m^T \mathbf \Psi}{\left((1+\kappa_D)\kappa_S\|\mathbf e_m^T\mathbf \Psi \boldsymbol {\tilde \theta}_0\|^2+\frac{(1+\kappa_D)\sigma_{n}^2}{\tilde P}\right)^2}+2(1+\kappa_D)\kappa_S\lambda_\text{max}(\mathbf \Omega) (\boldsymbol {\tilde \theta}_0^H-\boldsymbol {\tilde \theta}^H)
\end{align}
\setcounter{equation}{\value{mytempeqncnt17}}
\vspace*{4pt}
\hrulefill
\end{figure*}
\section{Proof of Theorem~\ref{theo:converge}}\label{app:theo}
Since problems \eqref{eq:SNRMaxProbwtheta3} and \eqref{eq:SNRMaxProbwtheta0} are equivalent, we focus on the convergence proof for problem \eqref{eq:SNRMaxProbwtheta3}. Define the objective function of problem \eqref{eq:SNRMaxProbwtheta3} and its lower bound in Proposition~\ref{prop:MMbnd} by $f(\boldsymbol {\tilde \theta})$ and $\bar f(\boldsymbol {\tilde \theta};\boldsymbol {\tilde \theta}_0)$, respectively. Denote the constraint set of problem \eqref{eq:SNRMaxProbwtheta3} by $\mathcal{S}$. Then, in order to prove Theorem~\ref{theo:converge}, we need to verify the following four conditions according to \cite[Section III]{Zhao2017TSP}:

1) $f(\boldsymbol {\tilde \theta}) \geq \bar f(\boldsymbol {\tilde \theta};\boldsymbol {\tilde \theta}_0),\ \forall \boldsymbol {\tilde \theta} \in \mathcal{S}$;

2) $f(\boldsymbol {\tilde \theta}_0) = \bar f(\boldsymbol {\tilde \theta}_0;\boldsymbol {\tilde \theta}_0)$;

3) $\nabla f(\boldsymbol {\tilde \theta}_0)=\nabla \bar f(\boldsymbol {\tilde \theta}_0;\boldsymbol {\tilde \theta}_0)$;

4) $\bar f(\boldsymbol {\tilde \theta};\boldsymbol {\tilde \theta}_0)$ is continuous in both $\boldsymbol {\tilde \theta}$ and $\boldsymbol {\tilde \theta}_0$.

\noindent Note that the convergence of the MM algorithm is guaranteed by the first two conditions while conditions 3) and 4) guarantee convergence to a stationary
point.

Clearly, conditions 1) and 4) hold due to Proposition~\ref{prop:MMbnd}. Moreover, from \eqref{eq:fbndfinal}, we have \eqref{eq:fbar},
where (a) holds since $\boldsymbol {\tilde \theta}_0 \in \mathcal{S}$, i.e., $|(\tilde \theta_0)_i|=1,\ i=1,\cdots,N_I+1$, and (b) is obtained based on the definition of $ \mathbf \Xi_0$ in Proposition~\ref{prop:MMbnd}. Therefore, condition 2) is also verified.

Concerning condition 3), we first rewrite $f(\boldsymbol {\tilde \theta})$ in \eqref{eq:objSDR} by
\setcounter{equation}{36}
\begin{align}
f(\boldsymbol {\tilde \theta})=\sum_{m=1}^{N_S} \frac{\|\mathbf e_m^T\mathbf \Psi \boldsymbol {\tilde \theta}\|^2}{(1+\kappa_D)\kappa_S\|\mathbf e_m^T\mathbf \Psi \boldsymbol {\tilde \theta}\|^2+\frac{(1+\kappa_D)\sigma_{n}^2}{\tilde P}}.
\end{align}
Then, we can readily calculate $\nabla f(\boldsymbol {\tilde \theta})$ by
\begin{align}\label{eq:nablaf}
&\nabla f(\boldsymbol {\tilde \theta})=\frac{2(1+\kappa_D)\sigma_{n}^2}{\tilde P}\nonumber \\ & \times \sum_{m=1}^{N_S} \frac{\boldsymbol {\tilde \theta}^H \mathbf \Psi^H  \mathbf e_m \mathbf e_m^T \mathbf \Psi}{\left((1+\kappa_D)\kappa_S\|\mathbf e_m^T\mathbf \Psi \boldsymbol {\tilde \theta}\|^2+\frac{(1+\kappa_D)\sigma_{n}^2}{\tilde P}\right)^2}.
\end{align}
On the other hand, since $\| \boldsymbol {\tilde \theta}\|^2=N_I+1$, we express $\bar f(\boldsymbol {\tilde \theta};\boldsymbol {\tilde \theta}_0)$ by
\begin{align}
& \bar f(\boldsymbol {\tilde \theta};\boldsymbol {\tilde \theta}_0)=\nonumber \\ &2\Re\{\boldsymbol {\tilde \theta}_0^H(\mathbf \Psi^H \mathbf \Xi_0^{-1}\mathbf \Psi -(1+\kappa_D)\kappa_S(\mathbf \Omega-\lambda_\text{max}(\mathbf \Omega)\mathbf I_{N_I+1})) \boldsymbol {\tilde \theta}\}\nonumber \\ &-(1+\kappa_D)\kappa_S\| \boldsymbol {\tilde \theta}\|^2\lambda_\text{max}(\mathbf \Omega) -(1+\kappa_D)\kappa_S(N_I+1)\lambda_\text{max}(\mathbf \Omega)\nonumber \\
&+\boldsymbol {\tilde \theta}_0^H(2(1+\kappa_D)\kappa_S\mathbf \Omega-\mathbf \Psi^H \mathbf \Xi_0^{-1}\mathbf \Psi)\boldsymbol {\tilde \theta}_0.
\end{align}
Accordingly, $\nabla \bar f(\boldsymbol {\tilde \theta};\boldsymbol {\tilde \theta}_0)$ is obtained by \eqref{eq:nablabarf},
where (a) and (b) are achieved based on the definitions of $\mathbf \Omega$ and $\mathbf \Xi_0$, respectively. By setting $\boldsymbol {\tilde \theta}=\boldsymbol {\tilde \theta}_0$ for $\nabla f(\boldsymbol {\tilde \theta})$ in \eqref{eq:nablaf} and $\nabla {\bar f}(\boldsymbol {\tilde \theta};\boldsymbol {\tilde \theta}_0)$ in \eqref{eq:nablabarf}, we eventually arrive at condition 3). Therefore, the proof is completed.

\end{appendices}

\end{document}